\documentclass[aps,prd,onecolumn,nofootinbib,superscriptaddress]{revtex4-2}
\usepackage{amssymb}
\usepackage{amsfonts}
\usepackage{amsmath}
\usepackage{amsfonts}
\usepackage{amsmath,amssymb}
\usepackage{epsfig,graphicx}
\usepackage{tikz}
\usepackage[compat=1.1.0]{tikz-feynman}
\usepackage{hyperref}
\hypersetup{hidelinks}

\begin{document}

\title{$SL(2N,C)$ Yang--Mills Theories: Direct Internal Forces and Emerging Gravity}

\author{J. L.~Chkareuli}
\affiliation{Institute of Theoretical Physics, Ilia State University, 0179 Tbilisi, Georgia}
\affiliation{Andronikashvili Institute of Physics, Tbilisi State University, 0177 Tbilisi, Georgia}

\begin{abstract}
A four-dimensional gauge-gravity unification based on local $SL(2N,C)$ symmetry is developed in a universal Yang--Mills-type setting, which, however, appears dynamically consistent only in the symmetry-broken phase. In the exact symmetry limit the theory may only be formulated in a premetric framework, where the accompanying tetrad multiplets, though promoted to dynamical fields, do not yet satisfy the conventional invertibility conditions. An ordinary Einstein--Cartan spacetime geometry emerges only in the broken post-soldering phase, in which the $SL(2N,C)$ tetrad multiplets are treated as constrained dynamical fields selecting a neutral internal symmetry branch. This realizes the breaking $SL(2N,C)\to SL(2,C)\times SU(N)$, thereby lifting all noncompact internal directions, while the surviving neutral tetrad is, as usual, associated with the gravitational field. A special ghost-free curvature-squared Lagrangian provides a consistent quadratic sector for the spin connection, propagating only admissible connection modes: the massless $SU(N)$ vector fields together with massive axial-vector and pseudoscalar multiplets. The Einstein--Cartan linear curvature term is argued to arise radiatively from fermion loops, thereby relating the gravitational scale to the same $SL(2N,C)$-covariant matter sector that defines the unified gauge coupling. Finally, the matter sector points to a deeper elementarity of $SL(2N,C)$ spinors, identified with preon constituents whose bound states form the observed quarks and leptons. Anomaly matching between preons and composites singles out $N=8$. The chain $SL(16,C)\to SL(2,C)\times SU(8)$ then naturally yields three composite quark--lepton families, while filtering out extraneous heavy states.
\end{abstract}

\maketitle

\section{Introduction}

It has long been recognized \cite{Utiyama, Kibble} that gravity can be
cast as a gauge theory in close analogy with the gauge description of
electromagnetic, weak, and strong interactions. In this language the spin
connection associated with local Lorentz transformations plays the role of a
gauge field, much as Yang--Mills bosons do for internal symmetry groups \cite{YM}.
Once this parallel is taken seriously, it is natural to ask whether
gravitational and internal gauge symmetries can be embedded into a single
non-compact group, so that all interactions are described within a common
gauge-theoretic framework. Such possibilities have been discussed in various
forms \cite{ish, ish1, cho, hu, per, cham, jpl}. However, despite their
attractive algebraic structure, these schemes face well-known obstacles,
which we consider here in the unified $SL(2N,C)$ symmetry case.

For the gravitational sector of such unification the relevant local symmetry
is $SL(2,C)$, the double cover of the proper orthochronous Lorentz group,
which furnishes the appropriate spinorial representation for fermions in
four-dimensional spacetime. If one assumes that fundamental spinor fields
are also responsible for organizing internal quantum numbers, it is natural
to extend $SL(2,C)$ to a larger group $SL(2N,C)$, in which $SU(N)$ appears
as an internal symmetry subgroup. In what follows we refer to such $SL(2N,C)$%
-based constructions as hyperunified theories (HUTs), since they combine $%
SL(2,C)$ gauge gravity and $SU(N)$ grand unification within a single local
group. The internal $SU(N)$ factor will be called hyperflavor, in order to
distinguish it from the Lorentzian $SL(2,C)$ sector, which remains neutral
under internal charges. The full $SL(2N,C)$ gauge multiplet contains vector,
axial-vector, and Lorentz-tensor connection fields, which provide a possible
algebraic framework for combining Standard Model interactions with gravity,
provided the known dynamical obstacles are resolved.

First, the unavoidable axial-vector and tensor partners of the usual vector
fields have no experimental support and therefore must either decouple or
lie at very high energies. Moreover, while internal forces are naturally
described by Yang--Mills kinetic terms quadratic in the field strength,
gravity in its Einstein--Cartan form is linear in curvature. A naive unified
Lagrangian therefore typically contains both linear and quadratic curvature
terms with different couplings. This tends to weaken the idea of a fully
unified gauge sector and may generically lead to multiple tensor
``graviton-like'' modes carrying $SU(N)$ quantum numbers, together with
ghost or tachyon modes in standard $R+R^2$ constructions.

A further difficulty is that the matter sector of a hyperunified $SL(2N,C)$
theory is strongly constrained. Higher non-fundamental representations of $%
SL(2N,C)$ generally mix fields of different spins inside a single multiplet
and therefore are not suitable for describing purely fermionic states such
as quarks and leptons. Conversely, the fundamental $2N$-dimensional
representation contains only spin-$1/2$ fields, but its internal $SU(N)$
quantum numbers do not match the observed pattern of the Standard Model
charges. Thus $SL(2N,C)$ cannot act directly on quarks and leptons without
introducing a large number of unobserved higher-spin or exotic-charged
states. This strongly suggests that quarks and leptons may not be truly
elementary in the $SL(2N,C)$ sense, but instead emerge as composites of more
fundamental preons. If these preons are chiral spin-$1/2$ fermions
transforming in the fundamental representation of $SL(2N,C)$, the group
becomes a natural candidate for organizing both gravitational and internal
dynamics at very short distances.

Finally, a conceptual issue concerns the Coleman--Mandula theorem \cite{18},
which severely restricts nontrivial unifications of spacetime and internal
symmetries within a simple Lie algebra. However, the $SL(2N,C)$ framework
developed here does not propose such a direct unification at the level of an
ordinary $S$-matrix theory on a fixed metric background. In the exact
symmetry limit, the theory is formulated on a premetric differentiable
manifold, while the usual Einstein--Cartan spacetime geometry and internal
symmetries emerge only in the spontaneously broken phase. This phase lifts
the noncompact directions through tetrad dynamics, leaving the residual
low-energy symmetry in the standard product form $SL(2,C)\times SU(N)$.

The central idea in this work is that the tetrad can simultaneously serve
several distinct roles. First, it is, as usual, the soldering tool which
connects the rigid spinorial tangent space with the flexible spacetime
manifold and thereby determines the true metric structure of spacetime.
Second, the tetrad is treated as a genuine dynamical field. Its kinetic
sector, which generally contains ghostlike components, is chosen to be the
Teleparallel-Gravity-type (TGR) torsion invariant \cite{tg}, thereby giving the
healthy Fierz--Pauli kinetic structure for the neutral graviton and
simultaneously supplying algebraic lifting terms for the broken connection
components. Third, a nonlinear $\sigma $-model-type constraint \cite{W} fixes a nonzero invariant soldering norm of the tetrad multiplets, while their hyperflavor orientation is selected dynamically by the Coleman--Weinberg effective potential. The resulting hyperflavor-blind vacuum spontaneously breaks $SL(2N,C)$ down to $SL(2,C)\times SU(N)$, leaving the neutral gravitational tetrad and the $SU(N)$ vector fields light, while the axial-vector and Lorentz-tensor submultiplets become heavy. Finally, in the presence of suitable fermion multiplets, the same
tetrad and tensor fields can radiatively generate an Einstein--Cartan (EC)
linear curvature term, even if such a term is absent at the tree level.

To make the quadratic-curvature sector consistent, we adopt a particular
ghost-free combination (GFC) of curvature squared terms originally
identified by Neville \cite{nev,nev1}, adapted here to the $SL(2N,C)$
symmetry case. We show that this structure arises naturally from an $%
SL(2N,C) $-invariant quadratic Lagrangian once the tetrad is restricted to
be neutral in hyperflavor space. With this choice, the field-strength-squared
Lagrangian becomes fully unified: a single gauge coupling governs the
dynamics of the vector, axial-vector, and tensor submultiplets.

Matter fields in the $SL(2N,C)$ framework are then reconsidered from a
composite perspective. We argue that a preon model with chiral fermions in
fundamental $SL(2N,C)$ representations, together with metacolor confinement
and anomaly matching, picks out $SL(16,C)$ with an $SU(8)$ hyperflavor
subgroup as a particularly compelling case. In this scenario, three chiral
quark-lepton families arise as composites of preons and fit naturally within
a single $SU(8)$-based structure.

Compared with the earlier formulations of $SL(2N,C)$ hyperunification \cite{ish1,hu,cham}, 
and particularly with the recent ones \cite{jpl}, the present work 
focuses on the dynamical origin of the broken post-soldering
phase. In the exact symmetry limit, the theory may only be formulated in a
premetric framework, where the tetrad multiplets, though promoted to
dynamical fields, do not yet satisfy the conventional invertibility
conditions. An ordinary EC spacetime geometry emerges only in the broken
post-soldering phase, in which the $SL(2N,C)$ tetrad multiplets are treated
as constrained dynamical fields selecting the neutral branch. The resulting
TGR tetrad kinetic structure both gives the healthy massless graviton and
lifts the broken noncompact directions, while the GFC curvature-squared
structure projects the spin-connection sector onto its ghost-free
propagating particle content. Another distinctive point is that, despite the
formal presence of the linear Einstein--Cartan term in the theory, this term
may be induced radiatively from the same $SL(2N,C)$-covariant matter
couplings, so that the gravitational scale need not be introduced as an
independent bare parameter, but may emerge universally from the unifying
Yang--Mills setting. Finally, the preon sector, already motivated in earlier
work \cite{jpl}, completes here an important matter-side interpretation of
the present dynamical construction: the true carriers of the full
hyperunified symmetry are not the observed quarks and leptons themselves,
but subelementary preon spinors whose bound states form the observed chiral
families.

The paper is organized as follows. Section 2 reviews the construction of $%
SL(2N,C)$ gauge theories, beginning with neutral spinors and $SL(2,C)$
symmetry, and then extending to charged spinors and total $SL(2N,C)$
framework. Section 3 is devoted to the role of tetrads in the theory being
crucial for the compatibility of spacetime geometry and internal symmetries.
Accordingly, the $SL(2N,C)$ theory is first formulated in a premetric
phase and then acquires EC spacetime in the broken post-soldering phase with
the residual $SL(2N,C)\rightarrow SL(2,C)\times SU(N)$ symmetry. Section 4
develops the quadratic strength-tensor Lagrangians and identifies a
ghost-free choice that treats all gauge submultiplets on an equal footing.
Section 5 explains how an EC gravity term is induced at one loop by fermion
bubbles involving tetrads and tensor fields. Section 6 reviews the full $%
SL(2N,C)$ hyperunification scenario in the spontaneously broken phase.\
Section 7 turns to the matter sector and shows how preon-based $SL(16,C)$
hyperunification can accommodate three composite quark-lepton families.
Section 8 summarizes the main results and outlines directions for further
work.

\section{$SL(2N,C)$ gauge theories}

Assuming that spinor fields provide the fundamental building blocks of both
spacetime and internal structure leads naturally from the $SL(2,C)$ gauge
group of gravity \cite{ish} to an enlarged local symmetry $SL(2N,C)$, where $%
N$ measures the size of an internal $SU(N)$ subgroup. This $SU(N)$ is
interpreted as a hyperflavor symmetry intended to organize the internal
quantum numbers of the quark-lepton sector, including color, weak isospin,
and family labels. In what follows, we outline how such a hyperflavor
symmetry arises and how it feeds into hyperunification.

\subsection{Neutral spinors: $SL(2,C)$ symmetry}

We start with the $SL(2,C)$ gauge symmetry case for a neutral spinor.
Consider a Dirac field $\Psi $ which transforms in the fundamental spinor
representation of $SL(2,C)$, acting in the tangent (spinor) bundle at each
spacetime point \cite{ish},
\begin{equation}
\Psi \rightarrow \Omega \Psi \text{, \ \ }\Omega =\exp \left\{ \frac{i}{4}%
\theta _{ab}\gamma ^{ab}\right\} \text{ }  \label{om}
\end{equation}%
where the matrix $\Omega $ satisfies the pseudounitarity condition
\begin{equation*}
\Omega ^{-1}=\gamma _{0}\Omega ^{+}\gamma _{0}\,
\end{equation*}

To preserve the invariance of the free kinetic term $i\overline{\Psi}
\gamma^\mu \partial_\mu \Psi$ under local $SL(2,C)$ transformations, the
constant matrices $\gamma^\mu$ must be replaced by tetrad-valued matrices $%
e^\mu$ that transform as
\begin{equation}
e^{\mu }\rightarrow \Omega e^{\mu }\Omega ^{-1}\text{ }  \label{trl}
\end{equation}
and are related to the usual tetrad fields $e^a{}_\mu$ by
\begin{equation*}
e^{\mu }=e_{a}^{\mu }\gamma ^{a}\text{ , \ }e_{\mu }=e_{\mu }^{a}\gamma _{a}
\text{ .}
\end{equation*}
Infinitesimally, this implies for the tetrad components
\begin{equation*}
\delta e^{\mu c}=\frac{1}{2}\theta _{ab}(e^{\mu a}\eta ^{bc}-e^{\mu b}\eta
^{ac})
\end{equation*}
while the usual invertibility conditions
\begin{equation}
e_{\mu }^{a}e_{a}^{\nu }=\delta _{\mu }^{\nu }\text{, \ }e_{\mu
}^{a}e_{b}^{\mu }= \delta _{b}^{a}  \label{or}
\end{equation}
ensure that these tetrads define a nondegenerate metric,
\begin{equation}
g_{\mu \nu }=\frac{1}{4}\mathrm{Tr}(e_{\mu }e_{\nu })=e_{\mu }^{a}e_{\nu
}^{b} \eta _{ab}\text{\ , \ }g^{\mu \nu }=\frac{1}{4}\mathrm{Tr}(e^{\mu
}e^{\nu })= e_{a}^{\mu }e_{b}^{\nu }\eta ^{ab}\,.  \label{gmn}
\end{equation}

Promoting $SL(2,C)$ to a local symmetry, $\theta _{ab}\rightarrow \theta
_{ab}(x)$, requires introducing a spin-connection one-form $I_{\mu }$
transforming as
\begin{equation*}
I_{\mu }\rightarrow \Omega I_{\mu }\Omega ^{-1}-\frac{1}{ig}(\partial _{\mu
}\Omega )\Omega ^{-1}
\end{equation*}%
and defining the covariant derivative
\begin{equation}
\partial _{\mu }\Psi \rightarrow D_{\mu }\Psi =\partial _{\mu }\Psi
+igI_{\mu }\Psi \text{ ,}  \label{124}
\end{equation}%
with gauge coupling $g$. The spin connection can be expanded in terms of
Dirac generators as
\begin{equation}
I_{\mu }=\frac{1}{4}T_{\mu }^{[ab]}\gamma _{ab}  \label{125}
\end{equation}%
where $T_{\mu }{}^{[ab]}$ transforms as
\begin{equation*}
\delta T_{\mu }^{[ab]}=\frac{1}{2}\theta _{\lbrack cd]}[(T_{\mu }^{[ac]}\eta
^{bd}-T_{\mu }^{[ad]}\eta ^{bc})-(T_{\mu }^{[bc]}\eta ^{ad}-T_{\mu
}^{[bd]}\eta ^{ac})]-\frac{1}{g}\partial _{\mu }\theta ^{\lbrack ab]}
\end{equation*}%
i.e. as a Lorentz-antisymmetric tensor in the local frame and a spacetime
one-form.

When squared with the appropriate tetrad contractions, this field strength
\begin{equation*}
T_{\mu \nu }^{[ab]}=\partial _{\lbrack \mu }T_{\nu ]}^{[ab]}+g\eta
_{cd}T_{[\mu }^{[ac]}T_{\nu ]}^{[bd]}\,
\end{equation*}%
leads to an $SL(2,C)$ invariant\ Yang-Mills type Lagrangian. However, in the
presence of a tetrad, an additional invariant term linear in $T_{\mu \nu
}^{[ab]}$, can also be written as
\begin{equation}
e\mathcal{L}_{G}^{(1)}=\frac{1}{2\kappa }e_{[a}^{\mu }e_{b]}^{\nu }T_{\mu
\nu }^{[ab]}\text{ , \ \ }e\equiv \lbrack -\det \mathrm{Tr}(e^{\mu }e^{\nu
})/4]^{-1/2}  \label{127}
\end{equation}%
where $\kappa $ is a dimensionful coefficient. Identifying $\kappa $ with $%
8\pi /M_{P}^{2}$ yields the EC action in Palatini formulation \cite{ish, heh}%
. Varying with respect to $T_{\mu }^{[ab]}$ expresses it algebraically in
terms of the tetrad and its derivatives, and one recovers the usual Einstein
action plus torsion-induced four-fermion terms \cite{heh}. The factor $e$ ensures
invariance under general four-dimensional diffeomorphisms as well, while the
$SL(2,C)$ gauge group solely acts in the tangent spinorial space \cite{ish}.

The coupling of the fermion to $I_{\mu }$ through (\ref{124}, \ref{125})
produces an additional interaction term
\begin{equation*}
e\mathcal{L}_{M}=-\frac{1}{2}g\epsilon ^{abcd}T_{\mu \lbrack ab]}e_{c}^{\mu }%
\overline{\Psi }\gamma _{d}\gamma _{5}\Psi
\end{equation*}%
which, after eliminating torsion, leads to an effective four-fermion contact
term
\begin{equation*}
\kappa \left( \overline{\Psi }\gamma _{d}\gamma ^{5}\Psi \right) (\overline{%
\Psi }\gamma ^{d}\gamma ^{5}\Psi )
\end{equation*}%
characteristic of EC gravity \cite{Kibble}.

We show below,\ in Sec. 5, that in the $SL(2N,C)$\ hyperunified theory,
whose fundamental bosonic sector is governed by a universal
quadratic-strength Lagrangian with a single gauge coupling, the EC term need
not be introduced as an independent tree-level input: it can instead be
generated radiatively by the corresponding fermion-loop diagrams.

\subsection{Charged spinors: $SL(2N,C)$ hyperunification}

We now pass to the full $SL(2N,C)$ group, which contains the spacetime
symmetry $SL(2,C)$ and an internal hyperflavor $SU(N)$ among its primary
subgroups. The $(8N^{2}-2)$ generators of $SL(2N,C)$ can be written as
tensor products of Dirac matrices with $SU(N)$ generators. A convenient
parametrization \cite{ish1} of a general transformation acting on fermions
is
\begin{equation}
\Omega =\exp \left\{ \frac{i}{2}\left[ \left( \theta ^{k}+\theta
_{5}^{k}\gamma _{5}\right) \lambda ^{k}+\frac{1}{2}\theta _{ab}^{K}\gamma
^{ab}\lambda ^{K}\right] \right\} \text{ \ \ }(K=0,k)  \label{rt}
\end{equation}%
where $\lambda ^{k}$ ($k=1,\dots ,N^{2}-1$) are the standard $SU(N)$
Gell-Mann matrices and $\lambda ^{0}=\sqrt{2/N}\,\mathbf{1}_{N}$ denotes the
normalized hyperflavor singlet matrix. The parameters $\theta ^{k}$, $\theta
_{5}^{k}$ and $\theta _{ab}^{K}$ may be global or local\footnote{%
The $\gamma $ and $\lambda $ matrices obey
\begin{eqnarray*}
\gamma ^{ab} &=&i[\gamma ^{a},\gamma ^{b}]/2,\text{ }\{\gamma ^{a},\gamma
^{b}\}=2\eta ^{ab}\mathbf{1},\text{ } \\
\lbrack \lambda ^{k},\lambda ^{l}] &=&2if^{klm}\lambda ^{m},\text{ }%
\{\lambda ^{k},\lambda ^{l}\}=\frac{4}{N}\delta ^{kl}\widehat{1}%
+2d^{klm}\lambda ^{m}\,.
\end{eqnarray*}%
}. Uppercase indices $I,J,K$ refer to the singlet-plus-adjoint $U(1)\times
SU(N)$ type basis, while lowercase indices $i,j,k$ denote the purely $SU(N)$
adjoint part. 

Equation~(\ref{rt}) describes the action of the $SL(2N,C)$ symmetry on
the fermion multiplet $\Psi$, namely, on Dirac spinors carrying an
additional hyperflavor $SU(N)$ index, so that $\Psi$ lives in the space 
$\mathbb{C}^4\otimes\mathbb{C}^N$. Equivalently, in the chiral
basis for $\gamma$ matrices, this symmetry acts separately
on the left- and right-handed Weyl components, $\Psi_{L}$ and $\Psi_{R}$, through the transformation matrices
\begin{eqnarray}
\Omega _{L} &=&\exp \left\{ \frac{i}{2}\left[ \left( \theta ^{k}-\theta
_{5}^{k}\right) \lambda ^{k}+\frac{1}{2}\theta _{ab}^{K}\Sigma
^{ab}\lambda ^{K}\right] \right\} ,\text{ \ }  \notag \\
\Omega _{R} &=&\exp \left\{ \frac{i}{2}\left[ \left( \theta ^{k}+\theta
_{5}^{k}\right) \lambda ^{k}+\frac{1}{2}\theta _{ab}^{K}\overline{%
\Sigma }^{ab}\lambda ^{K}\right] \right\} \text{ \ }  \label{lr}
\end{eqnarray}%
respectively. Here $\Sigma ^{ab}$ and $%
\overline{\Sigma }^{ab}$ are given by
\begin{equation*}
\Sigma ^{ab}=\frac{i}{2}(\sigma ^{a}\overline{\sigma }^{b}-\sigma ^{b}%
\overline{\sigma }^{a}),\text{ \ }\overline{\Sigma }^{ab}=\frac{i}{2}(%
\overline{\sigma }^{a}\sigma ^{b}-\overline{\sigma }^{b}\sigma ^{a})
\end{equation*}%
where the two-dimensional matrices $\sigma ^{a}$ and $\overline{\sigma }^{a}$
are expressed in terms of the unit and Pauli matrices as%
\begin{equation*}
\sigma ^{a}=(\widehat{1},\boldsymbol{\sigma }),\text{ }\overline{\sigma }%
^{a}=(\widehat{1},-\boldsymbol{\sigma })  
\end{equation*}%
Accordingly, the action of the $SL(2N,C)$ gauge multiplet on both
left- and right-handed Weyl sectors is explicitly specified.

The tetrad multiplet in the full $SL(2N,C)$ theory may in general contain
both vector and axial-vector Dirac components,
\begin{equation}
e_{\mu }=(e_{\mu }^{aK}\gamma _{a}+e_{\mu 5}^{aK}\gamma _{a}\gamma
_{5})\lambda ^{K},\text{ \ \ }K=0,1,\dots ,N^{2}-1  \label{lllll}
\end{equation}%
and transforms as in (\ref{trl}), where $\Omega $ is now defined by (\ref{rt}%
). One can simplify this structure and eliminate the axial-vector components
by introducing an auxiliary scalar multiplet,
\begin{equation*}
S=\exp \{i[(s^{k}+p^{k}\gamma _{5})\lambda ^{k}+T_{[ab]}\gamma ^{ab}\lambda
^{K}/2]\}
\end{equation*}%
that transforms as $S\rightarrow \Omega S$. By requiring the invariant
combination $S^{-1}e_{\mu }S$ to have zero projection onto the axial-vector
directions, we impose the constraint%
\begin{equation*}
\mathrm{Tr}[\gamma _{a}\gamma _{5}\lambda ^{K}(S^{-1}e_{\mu }S)]=0
\end{equation*}%
with the trace taken over both Dirac and hyperflavor indices. Subsequently,
selecting $S=1$ within the gauge orbit of the pair ($e_{\mu },S$) yields the
unitary-gauge representative of the covariantly constrained theory, in which
the tetrad reduces to a purely vectorial form \cite{ish1}. Finally, we adopt
the tetrad form
\begin{equation}
e_{\mu }=e_{\mu }^{aK}\gamma _{a}\lambda ^{K},\text{ }e=e_{a}^{\mu K}\gamma
^{a}\lambda ^{K}  \label{fs}
\end{equation}%
so that the neutral tetrad components in $e_{\mu }$ and $e^{\mu }$ are
identified with the pure $SL(2,C)$ tetrads, $e_{\mu }^{a0}=e_{\mu }^{a}$ and
$e_{a}^{\mu 0}=e_{a}^{\mu }$.

As in the previous $SL(2,C)$ case, one may formally write the tetrad
invertibility condition
\begin{equation*}
\frac{1}{32}\mathrm{Tr}(e_{\mu }e^{\nu })=\delta _{\mu }^{\nu },\text{ }%
e_{\mu }^{aK}e_{a}^{\nu K}=\delta _{\mu }^{\nu },\text{ }e_{\mu
}^{aK}e_{b}^{\mu K}=\delta _{b}^{a}
\end{equation*}%
and corresponding metric tensors
\begin{eqnarray*}
\frac{1}{8}\mathrm{Tr}(e_{\mu }e_{\nu }) &=&e_{\mu }^{aK}e_{\nu
a}^{K}=g_{\mu \nu }, \\
\frac{1}{8}\mathrm{Tr}(e^{\mu }e^{\nu }) &=&e_{a}^{\mu K}e^{\nu aK}=g^{\mu
\nu }
\end{eqnarray*}%
where we have used the trace conditions for $\gamma $ and $\lambda $
matrices.

Localizing the transformation (\ref{rt}) introduces a gauge field $I_{\mu }$
transforming as
\begin{equation}
I_{\mu }\rightarrow \Omega I_{\mu }\Omega ^{-1}-\frac{1}{ig}(\partial _{\mu
}\Omega )\Omega ^{-1}  \label{gg}
\end{equation}%
with field strength
\begin{equation}
I_{\mu \nu }=\partial _{\lbrack \mu }I_{\nu ]}+ig[I_{\mu },I_{\nu }]
\label{i}
\end{equation}%
and fermionic covariant derivative
\begin{equation*}
\partial _{\mu }\Psi \rightarrow D_{\mu }\Psi =\partial _{\mu }\Psi
+igI_{\mu }\Psi \,.
\end{equation*}%
The gauge connection $I_{\mu }$ can be decomposed into vector, axial, and
tensor parts
\begin{equation}
I_{\mu }=V_{\mu }+A_{\mu }+T_{\mu }=\frac{1}{2}\left( V_{\mu }^{k}+A_{\mu
}^{k}\gamma _{5}\right) \lambda ^{k}+\frac{1}{4}T_{\mu }^{[ab]K}\gamma
_{ab}\lambda ^{K}\text{ \ \ }(K=0,k)  \label{ggggg}
\end{equation}%
where each submultiplet carries the appropriate spacetime, Lorentz and
internal indices. The corresponding contribution to the field strength reads
as%
\begin{eqnarray}
I_{\mu \nu } &=&\frac{1}{2}\partial _{\lbrack \mu }\left( V^{k}+A^{k}\gamma
_{5}\right) _{\nu ]}\lambda ^{k}-\frac{1}{2}f^{ijk}g\left( V^{i}+A^{i}\gamma
_{5}\right) _{\mu }(V^{j}+A^{j}\gamma _{5})_{\nu }\lambda ^{k}  \notag \\
&&+\frac{1}{4}\left( \partial _{\lbrack \mu }T_{\nu ]}^{[ab]K}\gamma
_{ab}\lambda ^{K}+i\frac{g}{4}T_{\mu }^{[ab]K}T_{\nu }^{[a^{\prime
}b^{\prime }]K^{\prime }}[\lambda ^{K}\gamma _{ab},\lambda ^{K^{\prime
}}\gamma _{a^{\prime }b^{\prime }}]\right)  \label{f}
\end{eqnarray}%
apart from some unessential $V$--$T$ and $A$--$T$ mixed terms\footnote{%
The general field strength (\ref{i}) contains, in addition to the sectoral
vector/axial-vector and tensor contributions displayed in (\ref{f}), the
mixed $V$--$T$ and $A$--$T$ commutator terms%
\begin{equation*}
\Delta I_{\mu \nu }=-(g/4)f^{ijk}(V^{i}+A^{i}\gamma _{5})_{[\mu }T_{\nu
]}^{[ab]j}\gamma _{ab}\lambda ^{k}
\end{equation*}%
These terms describe the gauge interactions of the vector and axial-vector
submultiplets with the hyperflavored tensor fields. They introduce no
independent coupling and no new kinetic structure; after tetrad condensation
(presented below) they involve tensor fields belonging to the heavy sector.
For this reason they are suppressed in (\ref{f}), where only the sectoral
contributions relevant for constructing the independent quadratic-strength
Lagrangians are displayed.}. Meanwhile, the generic gauge-invariant matter
coupling can be written in terms of $e^{\mu }$ and the submultiplets as
\begin{equation}
e\mathcal{L}_{M}=-\frac{g}{2}\overline{\Psi }\left\{ e^{\mu },\left[ \frac{1%
}{2}\left( V_{\mu }^{k}+A_{\mu }^{k}\gamma _{5}\right) \lambda ^{k}+\frac{1}{%
4}T_{\mu }^{[ab]K}\gamma _{ab}\lambda ^{K}\right] \right\} \Psi \,.
\label{lfm}
\end{equation}%
Thus all three types of gauge fields couple with the same universal coupling
$g$ in the underlying $SL(2N,C)$ theory, which typically includes a standard
quadratic field-strength term for each submultiplet, while a possible linear
curvature term, analogous to (\ref{127}), may be induced radiatively (Sec.
5). The key question is how the observed low-energy spectrum can retain only
the $SU(N)$ vector fields and a single graviton, while the axial-vector and
tensor states are decoupled.

\section{Tetrads in $SL(2N,C)$ theory}

The full $SL(2N,C)$ gauge theory contains many more degrees of freedom than
are needed to describe known low-energy physics. Our strategy is to show that a dynamical tetrad, with its norm fixed by the length constraint and its hyperflavor orientation selected by quantum effects, can spontaneously break $SL(2N,C)$ and remove the unwanted non-compact directions. In this
picture the tetrad not only defines the spacetime geometry but also selects
the internal symmetry and filters the spin-connection multiplets that remain
light. The low-energy theory then exhibits only $SL(2,C)\times SU(N)$
symmetry with massless $SU(N)$ vectors, while hyperflavored axial-vector and
tensor fields are lifted to high masses. As will be seen below, a crucial
simplification occurs once tetrads are constrained to be neutral in
hyperflavor space. In that case only the $V_{\mu }^{k}$ and the singlet
tetrad $e_{\mu }^{a0}$ remain relevant for low-energy physics: the former
mediate internal forces, while the latter underlies the induced EC gravity.

\subsection{Gravity-inspired geometry}

The general tetrad ansatz (\ref{lllll}), even after elimination of the
axial-vector part, still includes the hyperflavor components as well (\ref%
{fs}). However, the invertibility conditions suitable for coupling to a
four-dimensional metric severely restrict this structure. Let us assume that
these conditions take the simple multiplicative form
\begin{equation}
e_{\mu }^{aK}e_{a}^{\nu K^{\prime }}=\delta _{\mu }^{\nu }n^{KK^{\prime }},%
\text{ }e_{\mu }^{aK}e_{b}^{\mu K^{\prime }}=\delta _{b}^{a}n^{KK^{\prime }}
\label{or1}
\end{equation}%
with some matrix $n$ whose form is to be determined. In the pure gravity
limit one has constraints
\begin{equation}
K=K^{\prime }=0;\text{ }n^{00}=1,\text{ }n^{kk}=0  \label{or1'}
\end{equation}%
which exclude a conventional $\delta ^{KK^{\prime }}$ form for the matrix $n$%
.

Multiplying (\ref{or1}) by $e_{\rho }^{bK^{\prime \prime }}$ and $%
e_{a}^{\rho K^{\prime \prime }}$ and using basic algebraic identities one
finds
\begin{equation*}
e_{\mu }^{aK}=n^{K0}e_{\mu }^{a0}\text{, \ }e_{a}^{\mu K^{\prime
}}=n^{K^{\prime }0}e_{a}^{\mu 0}
\end{equation*}%
which leads to the equality
\begin{equation*}
e_{\mu }^{aK}e_{a}^{\mu K^{\prime }}/\text{ }4=n^{K0}n^{K^{\prime }0}
\end{equation*}%
Assuming\ that $n^{KK^{\prime }}$ is a\ constant numerical matrix, one
easily finds that the only consistent solution compatible with the
conditions (\ref{or1}, \ref{or1'})
\begin{equation*}
n^{KK^{\prime }}=\delta ^{K0}\delta ^{K^{\prime }0}
\end{equation*}%
i.e.\ the standard invertibility relations hold only for the $K=0$
component. This implies that the general tetrad must effectively lie in the $%
SL(2,C)$ subgroup,
\begin{equation}
e_{\mu }^{aK}=e_{\mu }^{a}\delta ^{K0},\text{ }e_{a}^{\mu K}=e_{a}^{\mu
}\delta ^{K0}  \label{or6}
\end{equation}%
if it is to define a conventional spacetime metric. Thus, within the simple
factorized invertibility ansatz (\ref{or1}), the ordinary tetrad relations
select the hyperflavor-singlet direction. The neutral tetrad branch (\ref%
{or6}) should therefore be viewed not as an arbitrary truncation, but as the
branch on which the usual EC spacetime geometry is recovered.

One can enforce this by imposing (\ref{or6}) as a background constraint
which explicitly reduces the starting $SL(2N,C)$ to $SL(2,C)$ $\times $ $%
SU(N)$. Alternatively, when tetrads are treated as dynamical fields, the
neutral tetrad may emerge via the spontaneous breaking of $SL(2N,C)$, as we
see below.

\subsection{Premetric phase}

In general, the main difficulty is how to formulate the theory in a
manifestly $SL(2N,C)$-invariant way, since, as we have seen, a general
tetrad multiplet $e_{\mu }^{aK}$ does not satisfy the ordinary tetrad
invertibility condition in the unbroken phase. The compound index $(a,K)$
contains more components than the spacetime index $\mu $, so the full tetrad
multiplet cannot be treated as an ordinary square tetrad with a standard
inverse.

A useful way to formulate the theory before soldering is not to require the
matrix tetrad multiplet to possess an ordinary inverse. In the unbroken $%
SL(2N,C)$ phase one introduces two independent tetrad-type fields, the
"lower" and "upper" tetrads%
\begin{equation}
e_{\mu }=e_{\mu }^{aK}\gamma _{a}\lambda _{K},\qquad E^{\mu }=E_{a}^{\mu
K}\gamma ^{a}\lambda _{K},\qquad K=0,k  \label{Ee}
\end{equation}%
respectively, transforming in the adjoint representation of the local $%
SL(2N,C)$ group
\begin{equation*}
e_{\mu }\rightarrow \Omega e_{\mu }\Omega ^{-1},\qquad E^{\mu }\rightarrow
\Omega E^{\mu }\Omega ^{-1}
\end{equation*}%
In that invariant phase, the theory does not yet possess an ordinary metric
geometry. Instead, the lower and upper tetrad multiplets define two
independent invariant bilinears
\begin{equation*}
g_{\mu \nu }=e_{\mu }^{aK}e_{\nu }^{bK}\eta _{ab},\text{ }G^{\mu \nu
}=E_{aK}^{\mu }E_{bK}^{\nu }\eta ^{ab}
\end{equation*}%
which are not yet a metric and its inverse. These tensors can be used to
form invariant contractions of vectors and covectors, but in the unbroken
phase they are not inverse to one another. Thus the symmetric phase is best
understood as a pre-soldering bi-premetric spacetime, with no canonical
operation of raising and lowering indices and no ordinary EC metric
structure. Ordinary spacetime geometry appears only after the transition to the dynamically selected soldered neutral branch. On that
branch the tetrads reduce to the ordinary $SL(2,C)$ tetrad and inverse
tetrad, and the usual metric relation, $g_{\mu \nu }G^{\nu \rho }=\delta
_{\mu }^{\rho }$, emerges as a property of the broken phase rather than as
an assumption of the unbroken \ pre-soldering \ theory.

The same lower-upper pre-soldering formulation may be made also in the gauge
sector. The fields are still defined on a differentiable manifold with local
coordinates, ordinary partial derivatives, and local $SL(2N,C)$ gauge
transformations. What is absent in this phase is a prior physical metric
which would identify covariant and contravariant world indices.

Accordingly, besides the lower tetrad-type field $e_{\mu }$ and the upper
tetrad-type field $E^{\mu }$, one may introduce lower and upper gauge
connections, $I_{\mu }$ and $I^{\mu }$, and the corresponding covariant
derivatives
\begin{equation*}
D_{\mu }=\partial _{\mu }+igI_{\mu },\qquad D^{\mu }=\partial ^{\mu
}+igI^{\mu }.
\end{equation*}%
Here $\partial ^{\mu }$ is not obtained from $\partial _{\mu }$ by raising
an index with a background Minkowski metric. It is the contravariant
derivative operator of the premetric formulation and becomes $G^{\mu \nu
}\partial _{\nu }$ only in the soldered phase.

The $I_{\mu }$ and $I^{\mu }$ gauge multiplets and their strength-tensors $%
I_{\mu \nu }$ and $I^{\mu \nu }$are appropriately transformed under local $%
SL(2N,C)$ transformations (\ref{rt}, \ref{gg}), thus leading to the direct
invariant lower--upper pairing%
\begin{equation}
i_{0}=\mathrm{Tr}(I_{\mu \nu }I^{\mu \nu })  \label{i0}
\end{equation}%
The two premetrics $g_{\mu \nu }$ and $G^{\mu \nu }$ also allow the
lower-sector and upper-sector self-contractions
\begin{equation}
i_{1}=\mathrm{Tr}(I_{\mu \nu }I_{\rho \sigma })G^{\mu \rho }G^{\nu \sigma },%
\text{ \ \ }i_{2}=\mathrm{Tr}(I^{\mu \nu }I^{\rho \sigma })g_{\mu \rho
}g_{\nu \sigma }  \label{i12}
\end{equation}%
These are independent invariants in the pre-soldering phase, and if one
imposes the lower--upper exchange symmetry
\begin{equation*}
e_{\mu }\leftrightarrow E^{\mu },\qquad g_{\mu \nu }\leftrightarrow G^{\mu
\nu },\qquad I_{\mu }\leftrightarrow I^{\mu },
\end{equation*}%
then the two self-contractions must enter in the action with equal
coefficients. Thus the most general quadratic gauge sector may contain the
direct invariant and a symmetric pair of contracted invariants.

The fermion sector can be written in the same lower--upper symmetric form.
For a spinor multiplet $\Psi $,
\begin{equation*}
D_{\mu }\Psi =(\partial _{\mu }+igI_{\mu })\Psi ,\qquad D^{\mu }\Psi
=(\partial ^{\mu }+igI^{\mu })\Psi .
\end{equation*}%
The corresponding pre-soldering Dirac term may be written as
\begin{equation}
\mathcal{L}_{\Psi }=\frac{i}{4}\mathcal{E}\,\bar{\Psi}\left( E^{\mu }%
\overleftrightarrow{D}_{\mu }+e_{\mu }\overleftrightarrow{D}^{\mu }\right)
\Psi .  \label{psi}
\end{equation}%
Both pieces are $SL(2N,C)$ invariant: $e_{\mu }$ and $E^{\mu }$ transform
adjointly, while $\Psi $ transforms in the fundamental representation. The
factor $\mathcal{E}$ is the premetric volume density, reducing to the usual
tetrad determinant after soldering.

This formulation does not imply two independent gauge multiplets in the
low-energy theory. The soldering transition identifies the lower and upper
sectors. On the neutral tetrad branch (see below),
\begin{equation*}
g_{\mu \rho }G^{\rho \nu }=\delta _{\mu }^{\nu },\qquad E^{\mu }=G^{\mu \nu
}e_{\nu },
\end{equation*}%
and the upper connection is identified with the raised lower connection,
\begin{equation*}
I^{\mu }=G^{\mu \nu }I_{\nu },\qquad D^{\mu }=G^{\mu \nu }D_{\nu }.
\end{equation*}%
Consequently,
\begin{equation*}
I^{\mu \nu }=G^{\mu \rho }G^{\nu \sigma }I_{\rho \sigma },
\end{equation*}%
and all three quadratic gauge invariants $i_{0,1,2}$ in (\ref{i0}) and (\ref%
{i12}) reduce to the same ordinary Yang--Mills kinetic structure, differing
only by an overall normalization of the effective gauge coupling. Likewise,
the symmetric fermion term (\ref{psi}) reduces to the usual tetrad Dirac
action,
\begin{equation*}
e\mathcal{L}_{\Psi }\rightarrow \frac{i}{2}\,\bar{\Psi}e^{\mu }%
\overleftrightarrow{D}_{\mu }\Psi ,\text{\ }e\equiv \lbrack -\det \mathrm{Tr}%
(e^{\mu }e^{\nu })/4]^{-1/2}
\end{equation*}%
up to the conventional normalization. Thus the premetric theory is a
local $SL(2N,C)$-invariant premetric gauge theory, while the ordinary metric
Yang--Mills and fermion Lagrangians emerge only on the dynamically selected hyperflavor-blind soldered branch.\footnote{We recently became aware that a certain premetric approach to gauge-gravity unification was discussed earlier in the $SO(13,1)$ symmetry framework \cite{per}, 
which contains the local Lorentz generators and the $SO(10)$ GUT generators as 
commuting subalgebras. In that approach, the soldering one-form plays the role of an order parameter, 
while the unified-symmetry limit is associated with the vanishing of an ordinary metric. In contrast, 
the present construction uses the general $SL(2N,C)$ hyperunification framework with independent lower and upper premetrics, 
and the subsequent broken $SL(2N,C)\rightarrow SL(2,C)\times SU(N)$ post-soldering phase.}

\subsection{Towards tetrad condensation}

The pre-soldering $SL(2N,C)$-invariant theory contains two independent
tetrad-type fields (\ref{Ee}). The soldered phase leading to ordinary EC spacetime geometry is described by a sigma-model-type constraint, which in the simplest fixed-norm form can be presented as
\begin{equation}
\frac{1}{32}\mathrm{Tr}(e_{\mu }E^{\mu })=\frac{1}{4}e_{\mu
}^{aK}E_{aK}^{\mu }=1  \label{eql}
\end{equation}%
The tetrad multiplets themselves play the role of the soldering order parameter, while the constraint (\ref{eql}) fixes only its invariant magnitude and thus defines the nonlinear sigma-model surface of the broken phase. In the present effective description, the dynamics that produces the nonzero soldering norm is not explicitly modeled; the constraint (\ref{eql}) may be viewed as the fixed-norm limit of a more general tetrad potential. It does not select a definite $SU(N)$ hyperflavor direction. This orientation is instead selected dynamically by the Coleman--Weinberg effective potential considered below, whose preferred vacuum is the hyperflavor-blind branch. On this branch
we write
\begin{equation}
e_{\mu }^{aK}=\,e_{\mu }^{a}\delta ^{K0}+e_{\mu }^{\ast aK},\text{ }%
E_{a}^{\mu K}=e_{a}^{\mu }\delta ^{K0}+E_{a}^{\ast \mu K}  \label{par}
\end{equation}%
where, to maintain correspondence with the pure gravity case, we take
\begin{equation*}
e_{\mu }^{\ast a0}=0,\qquad E_{a}^{\ast \mu 0}=0
\end{equation*}%
(the star will denote hyperflavor-carrying quantities throughout). Once the
hyperflavor-blind branch (\ref{par}) is selected dynamically, the vacuum
realizes the spontaneous symmetry breaking of $SL(2N,C)$ down to $%
SL(2,C)\times SU(N)$. The neutral tetrad components $e_{\mu }^{a}$ and $%
e_{a}^{\mu }$, satisfying the standard invertibility relations (\ref{or}),
then define the soldered spacetime geometry, while the flavored components, $%
e_{\mu }^{\ast ak}$ and $E_{a}^{\ast \mu k}$, contain the modes associated
with the broken noncompact directions. Absorption of these modes induces
masses for the axial-vector and tensor field submultiplets, related to the $%
7(N^{2}-1)$ broken coset generators beyond the residual $SL(2,C)\times SU(N)$
symmetry.

With the parametrization (\ref{par}), the constraint (\ref{eql}) gives
\begin{equation}
\frac{1}{4}e_{\mu }^{aK}E_{a}^{\mu K}=1+e_{\mu }^{\ast ak}E_{a}^{\ast \mu
k}/4\rightarrow e_{\mu }^{\ast ak}E_{a}^{\ast \mu k}=0  \label{eE}
\end{equation}%
Thus the minimal scalar constraint removes only the total mixed trace of the
flavored sector. Meanwhile, the ordinary invertibility conditions
\begin{eqnarray*}
e_{\mu }^{aK}E_{a}^{\nu K} &=&\delta _{\mu }^{\nu }+e_{\mu }^{\ast
ak}E_{a}^{\ast \nu k}, \\
e_{\mu }^{aK}E_{b}^{\mu K} &=&\delta _{b}^{a}+e_{\mu }^{\ast ak}E_{b}^{\ast
\mu k}
\end{eqnarray*}%
hold only up to quadratic flavored corrections. Accordingly, the lower and
upper premetrics are

\begin{eqnarray*}
g_{\mu \nu } &=&e_{\mu }^{aK}e_{\nu }^{bK}\eta _{ab}=g_{\mu \nu }^{0}+g_{\mu
\nu }^{\ast }, \\
G^{\mu \nu } &=&E_{a}^{\mu K}E_{b}^{\nu K}\eta ^{ab}=g_{0}^{\mu \nu
}+G_{\ast }^{\mu \nu }
\end{eqnarray*}%
where $g_{\mu \nu }^{0}$ and $g_{0}^{\mu \nu }$ are the standard\ $SL(2,C)$
metric (\ref{gmn}), while%
\begin{equation*}
g_{\mu \nu }^{\ast }=e_{\mu }^{\ast ak}e_{\nu }^{\ast bk}\eta _{ab},\text{ }%
G_{\ast }^{\mu \nu }=E_{a}^{\ast \mu k}E_{b}^{\ast \nu k}\eta ^{ab}
\end{equation*}%
Thus, the product of full metrics
\begin{eqnarray*}
g_{\mu \rho }G^{\nu \rho } &=&\delta _{\mu }^{\nu }+g_{\mu \rho }^{0}G_{\ast
}^{\nu \rho }+g_{\mu \rho }^{\ast }g_{0}^{\nu \rho }+g_{\mu \rho }^{\ast
}G_{\ast }^{\nu \rho } \\
&=&\delta _{\mu }^{\nu }+O(e^{\ast }e^{\ast },E^{\ast }E^{\ast })
\end{eqnarray*}%
can significantly deviate from an exact EC geometry unless the flavored
metric corrections are removed or dynamically suppressed.

By itself, the $SL(2N,C)$ invariant constraint surface (\ref{eql}) possesses
a much larger accidental symmetry leaving many flat directions. To display
it, introduce the collective index
\begin{equation*}
Q=(\mu ,a,K),\qquad K=0,1,\ldots ,N^{2}-1,\qquad Q=1,\ldots ,16N^{2}
\end{equation*}%
and define $e_{Q}\equiv e_{\mu }^{aK}$ and $E_{Q}\equiv E_{a}^{\mu K}$. The
two premetric tetrads may then be assembled into a single complex vector
\begin{equation*}
\Phi ^{\mathcal{A}}=%
\begin{pmatrix}
e_{Q} \\[1mm]
E_{Q}%
\end{pmatrix}%
=%
\begin{pmatrix}
e_{\mu }^{aK} \\[1mm]
E_{a}^{\mu K}%
\end{pmatrix}%
,\qquad \mathcal{A}=1,\ldots ,32N^{2}.
\end{equation*}%
With the off-diagonal bilinear form
\begin{equation*}
\mathcal{J}=%
\begin{pmatrix}
0 & \mathbf{1}_{16N^{2}} \\
\mathbf{1}_{16N^{2}} & 0%
\end{pmatrix}%
,
\end{equation*}%
the singlet constraint becomes
\begin{equation*}
\frac{1}{8}\Phi ^{T}\mathcal{J}\Phi =1
\end{equation*}%
Therefore the isolated constraint possesses the accidental complex
orthogonal symmetry
\begin{equation}
SO(32N^{2},C),\qquad O^{T}\mathcal{J}O=\mathcal{J}  \label{32}
\end{equation}%
where $O$ is a general global rotation/boost in the full $32N^{2}$
dimensional complex tetrad space, mixing $e_{Q}$ and $E_{Q}$, while leaving
the singlet trace invariant. This symmetry appears because the constraint (%
\ref{eql}) does not resolve the separate geometric roles of the spacetime
index $\mu $, the Lorentz index $a$, and the hyperflavor index $K$; it only
sees the combined vector $\Phi $. In this representation the total premetric
tetrads $e_{\mu }^{aK}$ and $E_{a}^{\mu K}$ are the two isotropic halves of
the fundamental vector representation of $SO(32N^{2},C)$.

However, in the gravitational soldered phase, we considered above in (\ref%
{par}), the neutral tetrads $e_{\mu }^{a0}$ and $E_{a}^{\mu 0}$ are not
Goldstone fields of the symmetry breaking. They are retained as the real
neutral Higgs sector that defines the post-soldering EC spacetime geometry,
while the residual local Lorentz group $SL(2,C)$ remains unbroken.
Therefore, after the neutral tetrad sector saturates the constraint in (\ref%
{eE}), the relevant accidental symmetry for the remaining flat directions is
not the formal full symmetry (\ref{32}), but the flavored-sector symmetry $%
SO[32(N^{2}-1),C]$, which acts on the combined flavored tetrad vector $\Phi
^{\ast }=(e_{\mu }^{\ast ak},E_{a}^{\ast \mu k})^{T}$. The constraint (\ref%
{eql}) reduces to the condition $e_{\mu }^{\ast ak}E_{a}^{\ast \mu k}=0$ on
the charged tangent space, which is invariant under this accidental
symmetry. Thus the charged accidental zero-mode spectrum contains $%
32(N^{2}-1)$ independent flavored tetrad modes, whose number exactly
corresponds to the number of coset flat directions in the symmetry breaking
\begin{equation}
SO[32(N^{2}-1),C]\longrightarrow SO[32(N^{2}-1)-1,C]  \label{s0}
\end{equation}

This large accidental zero-mode spectrum must be distinguished from the true
gauge Goldstone spectrum of the underlying local theory, since this counting
refers to the accidental flat directions of the isolated soldering
constraint (\ref{eql}), not to the $7(N^{2}-1)$ Goldstone modes of the local
$SL(2N,C)$ gauge symmetry. They are removed in unitary gauge and become the
longitudinal components of the massive axial-vector and tensor field spin
connections. The remaining $25(N^{2}-1)$ charged zero modes are
pseudo-Goldstone modes emerging in the accidental symmetry breaking (\ref{s0}%
). Since the full theory possesses only the smaller local $SL(2N,C)$
symmetry, these modes are not protected and are expected to acquire large
masses from the remaining dynamics.

The flavored tetrads $e_{\mu }^{\ast ak}$ and $E_{a}^{\ast \mu k}$ are
generically massless fields. However, once the neutral soldering branch has
been selected, the co-tetrad fluctuation $E_{a}^{\ast \mu k}$ can be
converted to $E_{\mu }^{\ast ^{\prime }ak}=g_{\mu \nu }^{0}\eta
^{ab}E_{b}^{\ast \nu k}$, and the pseudo-Goldstone lifting appears as an
ordinary quadratic mass matrix for the pair ($e_{\mu }^{\ast ak},E_{\mu
}^{\ast ^{\prime }ak}$), summed over the adjoint index $k$. In
post-soldering phase, one may then impose, for simplicity, the extra $%
16(N^{2}-1)$ conditions, $e_{\mu }^{\ast ak}=E_{\mu }^{\ast ^{\prime }ak}$,
leaving in the theory a single tetrad multiplet $e_{\mu }^{aK}$ with the
flavored $e_{\mu }^{\ast ak}$ components. In the low-energy neutral branch
they may therefore be consistently neglected only after they are properly
lifted. Thus, the neutral tetrad presentation (\ref{or6}) should not be
interpreted as an explicit breaking of the underlying $SL(2N,C)$-invariant
formulation. It is the effective unitary-gauge representative of the
soldered branch after the gauge Goldstone modes have been absorbed and the
remaining hyperflavor-carrying tetrad modes have been lifted. Therefore,
both the pre-soldering covariant formulation and the broken post-soldering
phase can be described within an otherwise $SL(2N,C)$-invariant theory while
using the neutral tetrad branch for the low-energy EC spacetime geometry.
The full tetrad multiplet reduces to the neutral $SL(2,C)$ tetrad sector,
and all departures from it are either Goldstone modes eaten by the massive
spin connections or residual heavy flavored tetrads outside the leading
low-energy geometry. We will proceed with a neutral tetrad for the rest of
the analysis, rather than considering the general tetrad case first, and
subsequently substituting its vacuum-inspired form (\ref{or6}) into the
relevant equations.

\subsection{Dynamical tetrads: lifting noncompact directions}

We now promote the tetrad to a dynamical field. However, the naive quadratic
term
\begin{equation}
(D_{[\mu }e_{\nu ]}^{\;aK})(D^{[\mu }e^{\nu ]}{}_{aK})  \label{le}
\end{equation}%
is not appropriate as a fundamental tetrad kinetic term. A generic quadratic
tetrad-derivative form of this type propagates wrong-sign tetrad components.
The healthy gravitational kinetic term is instead provided by the
teleparallel equivalent of general relativity (TGR) \cite{tg}. In ordinary
TGR, the tetrad is used as the fundamental gravitational variable and the
Einstein--Hilbert scalar is replaced, up to a total derivative, by a special
quadratic combination of the torsion tensor.

In this case the tetrad torsion is
\begin{equation*}
C_{\mu \nu }{}^{a}=\partial _{\mu }e_{\nu }{}^{a}-\partial _{\nu }e_{\mu
}{}^{a},
\end{equation*}%
or, in local Lorentz components,
\begin{equation*}
C_{abc}=e_{b}{}^{\mu }e_{c}{}^{\nu }C_{\mu \nu a},\text{ }C_{abc}=-C_{acb},%
\text{ }C_{a}=C^{b}{}_{ba}
\end{equation*}%
So, the TGR torsion scalar is
\begin{equation*}
{\mathbb{T}}(C)=\frac{1}{4}C_{abc}C^{abc}+\frac{1}{2}%
C_{abc}C^{cab}-C_{a}C^{a}.
\end{equation*}%
Thus the healthy tetrad kinetic term is chosen as
\begin{equation}
e\mathcal{L}_{{\mathbb{T}}}=-M^{2}\,{\mathbb{T}}(C),  \label{tr}
\end{equation}%
where $M$ is the mass scale supplied by the TGR sector. Since the tetrad is
dimensionless, this is the only dimensionful parameter appearing in the
tetrad kinetic term.

The sign in (\ref{tr}) is fixed by the requirement that the quadratic
expansion around a nondegenerate tetrad background gives the correct
Fierz-Pauli kinetic term for the massless graviton. Writing
\begin{equation*}
g_{\mu \nu }=\eta _{\mu \nu }+h_{\mu \nu },\qquad h=\eta ^{\mu \nu }h_{\mu
\nu },
\end{equation*}%
one obtains, up to the overall normalization $M^{2}$,
\begin{equation}
e\mathcal{L}_{{\mathbb{T}}}(h)=\frac{M^{2}}{8}\left[ \partial _{\rho }h_{\mu
\nu }\partial ^{\rho }h^{\mu \nu }-2\partial _{\mu }h^{\mu \nu }\partial
^{\rho }h_{\rho \nu }+2\partial _{\mu }h^{\mu \nu }\partial _{\nu
}h-\partial _{\rho }h\,\partial ^{\rho }h\right]  \label{fp}
\end{equation}%
This is the standard Fierz-Pauli kinetic structure for a conventional
massless spin-two field. So, the tetrad kinetic term, being now the special
TGR combination, propagates the healthy graviton rather than the ghost modes
present in a generic quadratic tetrad derivative term (\ref{le}).

In the present $SL(2N,C)$ theory we keep the physical tetrad neutral or
hyperflavor-blind (\ref{or6}). We do not specialize $e_{\mu }^{a}{}$ to a
flat Minkowski background at this stage. It is the dynamical EC or TGR
tetrad defining the spacetime metric. The $SL(2N,C)$-covariant torsion is
obtained by replacing the ordinary derivative by the full covariant
derivative,
\begin{equation*}
C_{\mu \nu }=D_{\mu }e_{\nu }-D_{\nu }e_{\mu },
\end{equation*}%
where
\begin{equation*}
D_{\mu }e_{\nu }=\partial _{\mu }e_{\nu }+ig[I_{\mu },e_{\nu }],
\end{equation*}%
and $I_{\mu }$ is the full spin-connection multiplet (\ref{ggggg}). Thus,
\begin{equation}
C_{\mu \nu }=\partial _{\mu }e_{\nu }-\partial _{\nu }e_{\mu }+ig[I_{\mu
},e_{\nu }]-ig[I_{\nu },e_{\mu }]  \label{Cmunu}
\end{equation}%
For an axial-free tetrad $e_{\mu }=e_{\mu }^{aK}\gamma _{a}\lambda _{K}$,
the covariant torsion has the explicit Dirac decomposition
\begin{equation*}
C_{\mu \nu }=C_{\mu \nu }^{aK}\gamma _{a}\lambda _{K}+C_{\mu \nu
5}^{aK}\gamma _{a}\gamma _{5}\lambda _{K}
\end{equation*}%
There is no independent $\gamma _{ab}\lambda _{K}$ component in $C_{\mu \nu
} $. The reason is that the connection $I_{\mu }$ is Dirac-even, whereas the
tetrad is Dirac-odd; hence the commutator $[I_{\mu },e_{\nu }]$ is again
Dirac-odd. The tensor gauge fields $T_{\mu }^{[ab]K}$ are therefore not
absent; rather, they enter the vector and axial-vector coefficients $C_{\mu
\nu }^{aK}$ and $C_{\mu \nu 5}^{aK}$, but they do not generate a separate
Dirac-tensor torsion component proportional to $\gamma _{ab}\lambda _{K}$.

The $SL(2N,C)$ TGR Lagrangian has the same formal structure as (\ref{tr}),
but now the torsion components entering ${\mathbb{T}}(C)$ are built from the
flavored covariant object $C_{\mu \nu }$. In local indices,
\begin{equation*}
C_{abc}^{K}=e_{b}{}^{\mu }e_{c}{}^{\nu }C_{\mu \nu a}^{K},\qquad
C_{a}^{K}=C^{bK}{}_{ba},
\end{equation*}%
and the similar TGR combination is used:
\begin{equation}
{\mathbb{T}}(C)=\frac{1}{4}C_{abc}^{K}C^{abcK}+\frac{1}{2}%
C_{abc}^{K}C^{cabK}-C_{a}^{K}C^{aK}  \label{ltr1}
\end{equation}%
The normalization is chosen so that, for the neutral $K=0$ tetrad sector,
the action reduces to the ordinary TGR action (\ref{tr}).

The hyperflavor-blind vacuum branch is represented by the tetrad field configuration
\begin{equation}
\overline{e}_{\mu }=e_{\mu }{}^{a}\gamma _{a}\lambda ^{0},\qquad \lambda
^{0}=\sqrt{2/N}\,\mathbf{1}_{N}  \label{e}
\end{equation}%
providing the true vacuum in the theory.\footnote{Here ``vacuum'' refers only to 
the alignment of the tetrad multiplet in the neutral hyperflavor direction. 
The surviving field $e_\mu^{a}(x)$ remains the dynamical gravitational tetrad and is 
not fixed to a constant Minkowski value. The conventional flat-background choice 
$\langle e_\mu^{a}\rangle \sim \delta_\mu^{a}$ is a further specialization used only 
for a perturbative expansion around Minkowski spacetime.}
Since $\lambda ^{0}$ is proportional to the identity in hyperflavor space,
the vector generators commute with the tetrad field
configuration on this vacuum branch,
\begin{equation*}
\lbrack \lambda ^{k},\overline{e}_{\mu }]=0
\end{equation*}%
Consequently, the vector gauge fields $V_{\mu }^{k}$ do not enter the
algebraic part of the covariant torsion:
\begin{equation*}
ig[V_{\mu }^{k}\lambda _{k},\overline{e}_{\nu }]=0.
\end{equation*}%
The TGR term therefore gives no mass to the vector multiplet,
\begin{equation*}
m_{V}^{2}=0.
\end{equation*}%
Thus the $SU(N)$ vector gauge bosons remain massless at this stage.

The situation is different for the axial-vector generators. Their
contribution to the tetrad covariant derivative follows from
\begin{equation*}
ig\left[ \frac{1}{2}A_{\mu }^{k}\gamma _{5}\lambda _{k},e_{\nu }{}^{a}\gamma
_{a}\lambda ^{0}\right]
\end{equation*}%
Using the standard group condition for $\gamma $ and $\lambda $ matrices one
comes to
\begin{equation*}
D_{\mu }e_{\nu }\Big|_{A}=-ig\sqrt{2/N}\,e_{\nu }{}^{a}A_{\mu }^{k}\gamma
_{a}\gamma _{5}\lambda _{k},
\end{equation*}%
and hence
\begin{equation*}
C_{\mu \nu }\Big|_{A}=-ig\sqrt{2/N}\left( A_{\mu }^{k}e_{\nu }{}^{a}-A_{\nu
}^{k}e_{\mu }{}^{a}\right) \gamma _{a}\gamma _{5}\lambda _{k}.
\end{equation*}%
Thus the axial-vector fields enter the TGR torsion algebraically.
Substitution into the TGR scalar produces a mass term of the form, $%
m_{A}^{2}\sim g^{2}M^{2}$, once one uses the tetrad invertibility condition (%
\ref{or}). Note that the exact numerical coefficient depends only on the
normalization of the $SL(2N,C)$ generators: for the conventional
normalization taken in (\ref{ggggg}), $m_{A}^{2}=(36/N)g^{2}M^{2}$.

The tensor generators also do not commute with the tetrad. Their
contribution follows from
\begin{equation*}
ig\left[ \frac{1}{4}T_{\mu }^{[ab]K}\gamma _{ab}\lambda _{K},e_{\nu
}{}^{c}\gamma _{c}\lambda ^{0}\right]
\end{equation*}%
Using the Lorentz algebra relation
\begin{equation*}
\lbrack \gamma _{ab},\gamma _{c}]=2(\eta _{bc}\gamma _{a}-\eta _{ac}\gamma
_{b})
\end{equation*}%
one finds that the tensor connection contributes to the vector torsion
components as
\begin{equation}
C_{abc}^{K}(T)=g\sqrt{2/N}\left( T_{b\,[ac]}^{K}-T_{c\,[ab]}^{K}\right)
\label{CT}
\end{equation}%
up to the same generator-normalization convention. Therefore the TGR term (%
\ref{tr}, \ref{ltr1}) with these torsion components explicitly produces the
tensor-field mass block with a general factor $(2/N)g^{2}M^{2}$, which gives
mass of order $gM$ to the tensor spin-connection multiplet. The detailed
decomposition of $T_{a\,bc}^{K}$ into its irreducible Lorentz blocks will be
discussed later. At this stage it is enough to note that the tensor fields
are lifted by the same TGR structure that gives the healthy kinetic term for
the tetrad.

Thus, the TGR tetrad sector gives the desired pattern:
\begin{eqnarray}
V_{\mu }^{k} &:&\quad m_{V}^{2}=0,  \notag \\
A_{\mu }^{k} &:&\quad m_{A}^{2}\sim g^{2}M^{2},  \notag \\
T_{\mu }^{[ab]K} &:&\quad m_{T}^{2}\sim g^{2}M^{2}  \label{vat}
\end{eqnarray}%
This is precisely the pattern expected from the hyperflavor-blind tetrad.
The unbroken gauge sector is the vector $SU(N)$, while the non-compact
axial-vector and tensor directions are lifted. In group-theoretic terms, the
hyperflavor-blind tetrad selects the residual structure, $SL(2,C)\times
SU(N) $, with the Lorentz sector realized through the physical tetrad and
the $SU(N)$ vector bosons remaining massless.

It is important that the same TGR sign is responsible for both effects. It
gives the correct kinetic term for the dimensionless tetrad and hence for
the emergent graviton, while also giving the algebraic mass terms for the
non-compact spin-connection fields. In particular, as will be shown below,
this sign is also compatible with the healthy propagation of the
longitudinal pseudoscalar mode contained in the tensor multiplet. Thus the
TGR replacement of the naive kinetic terms (\ref{le}) is not merely a
cosmetic change; it is required both for the absence of tetrad ghosts and
for the correct lifting of the non-compact $SL(2N,C)$ directions.

Consequently, the theory contains the characteristic scale $M$, introduced
by the TGR tetrad sector. Above the scale $gM$, the full spin-connection
multiplet propagates according to the Yang--Mills-type curvature terms,
which we consider in the next section. Below this scale, the heavy
axial-vector and tensor connections can be integrated out. Their equations
of motion enforce the covariant tetrad compatibility condition in the
non-compact directions,
\begin{equation*}
D_{[\mu }e_{\nu ]}=0
\end{equation*}%
up to corrections suppressed by powers of $p^{2}/(g^{2}M^{2})$. The vector
gauge fields $V_{\mu }^{k}$ do not participate in this algebraic
compatibility condition for the neutral tetrad, because they commute with $%
\lambda ^{0}$ and hence with $\overline{e}_{\mu }$. They remain as massless $%
SU(N)$ gauge fields with their own Yang--Mills dynamics. Thus, at low
energies, integrating out the heavy axial-vector and tensor connections
leaves ordinary TGR, or equivalently Einstein gravity, for the neutral
tetrad, together with the massless $SU(N)$ vector sector and
higher-derivative corrections generated by the heavy fields.

The $SL(2,C)$ EC term (\ref{127} ) and the TGR tetrad-torsion term (\ref{tr}%
) are both dimension-two gravitational operators, but they play different
roles in the present construction. The EC term fixes the observed Planck
scale and dominates the infrared graviton dynamics when $M\ll M_{P}$. The
TGR term is introduced instead of the naive tetrad kinetic term (\ref{le})
because its specific torsion-scalar combination provides a ghost-free
kinetic term for the dynamical tetrad. Expanded around a nondegenerate
background, it gives the standard Fierz-Pauli kinetic structure (\ref{fp})
for the massless graviton rather than the ghost modes generated by a generic
quadratic tetrad-derivative term. At the same time, when this TGR torsion is
built with the full $SL(2N,C)$ covariant derivative, it provides algebraic
lifting of the non-compact axial-vector and tensor spin connections. Thus $M$
is the scale of tetrad-induced $SL(2N,C)$ breaking and of the heavy
connection masses, while $M_{P}$ continues to control Newton's constant.

The constraint (\ref{eql}) fixes only the overall length of the unified
tetrad, leaving a continuous set of possible orientations in hyperflavor
space. Moreover, in the generic parameterization (\ref{par}), part of the
flavored tetrad modes are absorbed by the broken tensor and axial-vector
gauge fields, while the remaining modes stay as physical fields and must be
given large masses. Both issues are resolved once quantum fluctuations of
the gauge fields are included: they generate a Coleman--Weinberg potential
\cite{cw} that selects the hyperflavor-blind vacuum and lifts the
hyperflavored symmetric modes.

\section{Lagrangians}

Having formulated the $SL(2N,C)$ invariant pre-soldering theory, we now turn
to the gauge-field Lagrangians in the post-soldering phase. Once the neutral
tetrad branch is selected, the two premetrics become inverse to one another,
and the upper gauge connection is identified with the raised lower
connection. Consequently the pre-soldering gauge invariants (\ref{i0}, \ref%
{i12}) reduce to the usual Yang--Mills type quadratic structures.

In general, an $SL(2N,C)$ invariant action may contain both linear and
quadratic terms in the field strengths, each with its own coupling. For our
purposes we focus on the quadratic sector and require it to be governed by a
single universal gauge coupling, common to vector, axial, and tensor parts
of the connection (\ref{ggggg}). The corresponding linear curvature term is
the Einstein--Cartan type action (\ref{127}). In the present approach it may
be viewed as an emergent contribution generated primarily by fermion loops.
In this way one can maintain hyperunification in the quadratic sector while
letting gravity arise from quantum effects, as will be seen in the next
section. The present section is therefore devoted mainly to the quadratic
strength-tensor Lagrangians, which determine the propagating gauge-field
content of the post-soldering theory.

We first consider the vector and axial-vector components of the connection
and then turn to the full gauge multiplet, where the tensor connection
requires the ghost-free curvature-squared combination.

\subsection{Vector and axial-vector fields}

We first examine the part of the post-soldering Lagrangian associated with
the $SU(N)$ vector and axial-vector fields, which are the basic spin-1
carriers of hyperflavor. From the quadratic strength term built out of (\ref%
{f}) one finds
\begin{eqnarray}
e\mathcal{L}_{H}^{VA} &=&-\frac{1}{2}\mathrm{Tr}[(V_{\mu \nu }^{k}\lambda
^{k}+\gamma _{5}A_{\mu \nu }^{k}\lambda ^{k})^{2}/4]=-\frac{1}{4}(V_{\mu \nu
}^{k})^{2}-\frac{1}{4}(A_{\mu \nu }^{k})^{2}  \label{I} \\
&=&-\frac{1}{4}[\partial _{\lbrack \mu }V_{\nu ]}^{k}-gf^{ijk}(V_{\mu
}^{i}V_{\nu }^{j}+A_{\mu }^{i}A_{\nu }^{j})]^{2}-\frac{1}{4}[\partial
_{\lbrack \mu }A_{\nu ]}^{k}-gf^{ijk}(V_{\mu }^{i}A_{\nu }^{j})]^{2}  \notag
\end{eqnarray}%
where $V_{\mu \nu }^{k}$ and $A_{\mu \nu }^{k}$ are the corresponding field
strengths.\ The vector fields thus have the standard $SU(N)$ gauge
structure. The axial-vector fields, however, are not gauge bosons of an
independent surviving axial gauge symmetry. They transform as adjoint fields
under the residual compact $SU(N)$, with interactions fixed by the original $%
SL(2N,C)$ curvature. The terms involving $A_{\mu }^{k}$ reflect the
noncompact part of the hyperunified algebra. After tetrad condensation these
axial-vector fields acquire mass (\ref{vat}) from the tetrad kinetic sector
and decouple at energies well below the scale $gM$.

In the matter sector, the vector fields couple to ordinary fermions through
\begin{equation*}
e\mathcal{L}_{HM}^{V}=-\frac{g}{2}V_{\mu }^{i}\overline{\Psi }e_{a}^{\mu
}\gamma ^{a}\lambda ^{i}\Psi
\end{equation*}%
whereas the axial-vector couplings vanish for a neutral tetrad, so $A_{\mu
}^{k}$ have no direct interactions with fermionic matter on the neutral
branch. They can still participate in loop processes through their
self-interactions and through their interactions with the vector fields.
Present experimental data \cite{pdg} do not forbid heavy states with such
quantum numbers in the multi-TeV range. Once they are integrated out, the
vector sector recovers gauge invariance at lower energies. Thus the
post-soldering vector sector contains massless $SU(N)$ gauge bosons $V_{\mu
}^{k}$, while the axial-vector fields $A_{\mu }^{k}$ are heavy adjoint
states associated with the lifted noncompact directions.

\subsection{Total gauge multiplet}

As discussed earlier, the quadratic part of a consistent post-soldering $%
SL(2N,C)$ theory must contain strength-tensor invariants for the vector,
axial-vector and tensor components of the total connection. For the vector
and axial-vector sectors the Yang--Mills form is sufficient. The tensor
sector is more restrictive: a generic curvature-squared spin-connection term
propagates ghosts. Among the possible combinations, the ghost-free
combination (GFC), identified for the first time in Refs.~\cite{nev,nev1}, stands out as the appropriate choice. In the pure
$SL(2,C)$ case, this Lagrangian reads
\begin{equation*}
e\mathcal{L}_{G}^{(2)}=\lambda T_{[ab]}\left(
T^{abcd}-4T^{acbd}+T^{cdab}\right)
\end{equation*}%
where $T^{abcd}=T_{\mu \nu }^{[ab]}e^{\mu c}e^{\nu d}$ and $T_{\mu \nu
}^{[ab]}{}$ is the curvature of the spin connection. Writing everything in
terms of the tensor fields and tetrads gives
\begin{equation}
e\mathcal{L}_{G}^{(2)}=\lambda (T_{[ab]}^{\mu \nu }T_{\mu \nu
}^{[ab]}-4T_{[ab]}^{\mu \nu }T_{\rho \nu }^{[ac]}e_{\mu c}e^{\rho
b}+T_{[ab]}^{\mu \nu }T_{\rho \sigma }^{[cd]}e_{\mu c}e_{\nu d}e^{\rho
a}e^{\sigma b})  \label{re}
\end{equation}%
where $\lambda $ is a constant. In this model the spin connection becomes a
propagating field, but not as an ordinary Yang--Mills gauge field. The
special relative coefficients in the GFC eliminate the dangerous ghost modes
and isolate the healthy spin-connection excitation identified as the massive
pseudoscalar state. In the present hyperunified theory the actual
propagating tensor-sector mode will be determined below after the TGR mass
lifting terms are included.

We now embed this structure into the post-soldering hyperunified $SL(2N,C)$
theory by writing a general quadratic Lagrangian built from the total
curvature and the neutral tetrad background
\begin{equation}
e\mathcal{L}_{H}^{(2)}=\lambda _{I}\mathrm{Tr}(aI^{\mu \nu }I_{\mu \nu
}+bI^{\mu \nu }I_{\rho \nu }e_{\mu }e^{\rho }+cI^{\mu \nu }I_{\rho \sigma
}e_{\mu }e_{\nu }e^{\rho }e^{\sigma })  \label{r1'}
\end{equation}%
with constants $a,b,c$ to be fixed by the requirement that the $SL(2,C)$
limit of (\ref{r1'}) reproduces (\ref{re}). It is important to distinguish
the origin of the ghost-free kinetic projector from the origin of the
algebraic lifting. The special GFC coefficients $(1,-4,1)$ are fixed solely
by the Lorentz spin-connection kinetic problem and therefore apply
independently to each hyperflavor component $T_{\mu}^{[ab]K}$. In the pure $%
SL(2,C)$ case the degeneracy of this projector is lifted by the linear
Einstein--Cartan term, whereas in the flavored $SL(2N,C)$ sectors no
corresponding EC term is available; there the same degeneracy is lifted
instead by the TGR-induced tensor mass term. Thus the GFC structure is
universal, while the lifting mechanism is sector-dependent.

Using neutral tetrads satisfying (\ref{or6}) and the trace properties of
Dirac and $SU(N)$ matrices, one finds for the vector and axial-vector
contributions
\begin{eqnarray*}
\mathrm{Tr}\left( I^{(W)\mu \nu }I_{\mu \nu }^{(W)}\right) &=&(W_{\mu \nu
}^{k}W^{\mu \nu k})/2,\text{ } \\
\mathrm{Tr}\left( I^{(W)\mu \nu }I_{\rho \nu }^{(W)}e_{\mu }e^{\rho }\right)
&=&2(W_{\mu \nu }^{k}W^{\mu \nu k}), \\
\mathrm{Tr}\left( I^{(W)\mu \nu }I_{\rho \sigma }^{(W)}e_{\mu }e_{\nu
}e^{\rho }e^{\sigma }\right) &=&4(W_{\mu \nu }^{k}W^{\mu \nu k})\text{ \ }%
(W\equiv V,A)
\end{eqnarray*}%
and for the tensor submultiplet
\begin{eqnarray*}
\mathrm{Tr}\left( I^{(T)\mu \nu }I_{\mu \nu }^{(T)}\right) &=&T_{[ab]}^{\mu
\nu K}T_{\mu \nu }^{[ab]K},\text{ } \\
\mathrm{Tr}\left( I^{(T)\mu \nu }I_{\rho \nu }^{(T)}e_{\mu }e^{\rho }\right)
&=&3T_{[ab]}^{\mu \nu K}T_{\mu \nu }^{[ab]K}+4(T_{[ab]}^{\mu \nu K}T_{\rho
\nu }^{[ac]K}e_{\mu c}e^{\rho b}), \\
\mathrm{Tr}\left( I^{(T)\mu \nu }I_{\rho \sigma }^{(T)}e_{\mu }e_{\nu
}e^{\rho }e^{\sigma }\right) &=&2T_{[ab]}^{\mu \nu K}T_{\mu \nu
}^{[ab]K}+8(T_{[ab]}^{\mu \nu K}T_{\rho \sigma }^{[cd]K}e_{\mu c}e_{\nu
d}e^{\rho a}e^{\sigma b})\text{ .}
\end{eqnarray*}%
Combining these pieces, the quadratic Lagrangian involves
\begin{equation}
(a/2+2b+4c)W^{2}+(a+3b+2c)TT+4b\,TTee+8c\,TTeeee  \label{W2}
\end{equation}%
where $W^{2}$ denotes the vector/axial-vector term and the symbols $TT$
etc.\ represent the tensor structures in (\ref{re}).

To reproduce the pure-gravity GFC combination (\ref{re}) one must impose
\begin{equation*}
a+3b+2c=1,\text{ }b=-1,\text{ }c=1/8\rightarrow a=15/4
\end{equation*}%
which fixes $a,b,c$ and yields a coefficient $(3/8)$ for $W^{2}$ in (\ref{W2}
). With these values the hyperunified quadratic Lagrangian becomes
\begin{eqnarray}
e\mathcal{L}_{H}^{(2)} &=&-\frac{1}{4}(V_{\mu \nu }^{k}V^{\mu \nu k}+A_{\mu
\nu }^{k}A^{\mu \nu k})  \label{RR} \\
&&-\frac{1}{4}\left( T_{[ab]}^{\mu \nu K}T_{\mu \nu }^{[ab]K}-4T_{[ab]}^{\mu
\nu K}T_{\rho \nu }^{[ac]K}e_{\mu c}e^{\rho b}+T_{[ab]}^{\mu \nu K}T_{\rho
\sigma }^{[cd]K}e_{\mu c}e_{\nu d}e^{\rho a}e^{\sigma b}\right)  \notag
\end{eqnarray}%
after choosing $\lambda _{I}=-2/3,$ so that the vector and axial sectors,
given above in (\ref{I}), have canonical normalization. A further rescaling
of tensor fields and of the coupling,
\begin{equation*}
T_{[ab]}^{\mu \nu K}\rightarrow \sqrt{8/3}T_{[ab]}^{\mu \nu K},\text{ }%
g\rightarrow \sqrt{3/8}g
\end{equation*}%
puts the tensor sector in canonical form as well. The result is a unified
quadratic post-soldering Lagrangian in which the vector, axial-vector and
tensor curvatures are normalized consistently, and a single gauge coupling $%
g $ controls the interactions of all components of the $SL(2N,C)$ connection.

\subsection{Propagating fields}

Whereas the massless vector and massive axial-vector submultiplets of the
total gauge multiplet $I_{\mu }$ (\ref{ggggg}) freely propagate, the tensor
field submultiplet may in general contain the ghostly kinetic terms. In this
connection, let us now examine which states are actually propagated by the
tensor part of the GFC term in the Lagrangian (\ref{RR}) and how they are
compatible with the lifting of non-compact directions mentioned above.

For this purpose it is sufficient to work in the quadratic approximation
around the neutral tetrad background and to keep the bilinear derivative
part of the tensor curvature. We write
\begin{equation*}
T_{\mu \nu }^{[ab]K}=\partial _{\mu }T_{\nu }^{[ab]K}-\partial _{\nu }T_{\mu
}^{[ab]K}
\end{equation*}%
and introduce local Lorentz indices,
\begin{equation*}
T_{abc}^{K}=e_{a}^{\mu }T_{\mu bc}^{K},\qquad T_{abc}^{K}=-T_{acb}^{K}
\end{equation*}%
For every hyperflavor index $K=0,1,\ldots ,N^{2}-1$, this can be decomposed
into irreducible Lorentz pieces as
\begin{equation*}
T_{abc}^{K}=q_{abc}^{K}+\frac{1}{3}\left( \eta _{ab}v_{c}^{K}-\eta
_{ac}v_{b}^{K}\right) +\frac{1}{6}\epsilon _{abcd}B^{dK}
\end{equation*}%
Here
\begin{equation*}
v_{a}^{K}=T_{ba}^{bK}{},\qquad B^{aK}=\epsilon ^{abcd}T_{bcd}^{K},
\end{equation*}%
while the remaining tensor $q_{abc}^{K}$ satisfies
\begin{equation*}
q_{ba}^{bK}=0,\qquad \epsilon ^{abcd}q_{bcd}^{K}=0.
\end{equation*}

The GFC in the Lagrangian (\ref{RR}) in local Lorentz indices looks as
\begin{equation}
\mathcal{N}%
_{T}^{K}=T_{abcd}^{K}T^{abcdK}-4T_{abcd}^{K}T^{acbdK}+T_{abcd}^{K}T^{cdabK}
\label{nk}
\end{equation}%
Substituting the axial irreducible component
\begin{equation*}
T_{abc}^{K}\Big|_{B}=\frac{1}{6}\epsilon _{abcd}B^{dK}
\end{equation*}%
into the GFC bracket (\ref{nk}), one obtains
\begin{equation*}
e\mathcal{L}_{T}^{(2)}\Big|_{B}=\frac{1}{4}\left( \partial
_{a}B^{aK}\right) ^{2}
\end{equation*}%
Thus the GFC term does not give a Maxwell-type kinetic term for the full
pseudo-vector $B_{a}^{K}$. It gives only a kinetic structure for its
divergence. The transverse components of $B_{a}^{K}$, as well as the $%
q_{abc}^{K}$ and $v_{a}^{K}$ irreducible components, do not propagate from
the GFC bracket alone. The possible propagating tensor-sector state is
therefore the longitudinal pseudoscalar combination
\begin{equation*}
\pi ^{K}\sim \partial _{a}B^{aK}.
\end{equation*}

This conclusion must be combined with the algebraic mass lifting terms
generated by the TGR tetrad sector considered in Section 3.4. In the
flavor-blind tetrad background (\ref{or6}) and with the standard
normalization of the $SL(2N,C)$ generators taken in (\ref{ggggg}), the
tensor contribution to the covariant torsion is given in (\ref{CT}). For the
axial irreducible tensor component this gives
\begin{equation*}
C_{abc}^{K}\Big|_{B}=-\frac{g}{3}\sqrt{\frac{2}{N}}\,\epsilon _{abcd}B^{dK}
\end{equation*}%
Due to its total antisymmetry, the trace vector $C_{a}^{K}\equiv C_{ba}^{bK}$
vanishes, so the last term in TGR torsion scalar ${\mathbb{T}}(C)$ does not
contribute to the $B_{a}^{K}$ sector. As a result, one obtains
\begin{equation*}
C_{abc}^{K}C^{abcK}=-\frac{4}{3N}g^{2}B_{a}^{K}B^{aK}
\end{equation*}%
and eventually
\begin{equation*}
{\mathbb{T}}(C)\Big|_{B}=\frac{1}{3N}g^{2}B_{a}^{K}B^{aK}
\end{equation*}

Combining the GFC derivative term with the TGR lifting term gives
\begin{equation}
e\mathcal{L}_{B}^{(2)}=\frac{1}{4}\left( \partial _{a}B^{aK}\right) ^{2}+%
\frac{1}{2}C_{B}B_{a}^{K}B^{aK},\text{ }C_{B}=-\frac{2}{3N}g^{2}M^{2}
\label{lb}
\end{equation}%
The sign of the algebraic term would be problematic if $B_{a}^{K}$ were an
ordinary propagating pseudovector. However, the GFC bracket does not
propagate $B_{a}^{K}$ as a full vector field. Only its longitudinal
pseudoscalar component, $\pi ^{K}=\partial _{a}B^{aK}$, is dynamical.
Varying (\ref{lb}) with respect to $B_{a}^{K}$ and taking the divergence
gives%
\begin{equation*}
(\square -2C_{B})\pi ^{K}=0
\end{equation*}
which can equivalently be represented by the effective Lagrangian
\begin{equation*}
e\mathcal{L}_{\mathrm{eff}}^{(\pi )}=\frac{1}{2}(\partial _{a}\pi
_{c}^{K})(\partial ^{a}\pi _{c}^{K})-\frac{1}{2}m_{\pi }^{2}\pi _{c}^{K}\pi
_{c}^{K},\text{ }\pi _{c}^{K}=\sqrt{3N/2}\,\pi ^{K}/gM
\end{equation*}%
for the canonically normalized pseudoscalar fields.

Remarkably, the sign fixed by the ordinary TGR sector in (\ref{tr}), which
provides the correct kinetic term for the tetrad and eventually for the
graviton, is precisely the sign required for the longitudinal pseudoscalar
selected by the GFC kinetic structure to be ghost-free and non-tachyonic.
The transverse components of $B_{a}^{K}$, together with $v_{a}^{K}$ and $%
q_{abc}^{K}$, remain non-propagating auxiliary fields at the quadratic
level. Hence the tensor connection does not propagate as a full tensor gauge
multiplet; it propagates one pseudoscalar state $\pi ^{K}$ for each
hyperflavor index $K$.

Thus, at the quadratic level, the propagating physical sector contained in (%
\ref{RR}), supplemented by the TGR lifting masses, may be written as
\begin{eqnarray}
e\mathcal{L}_{H\mathrm{phys}}^{(2)} &=&-\frac{1}{4}V_{\mu \nu }^{k}V^{\mu
\nu k}-\frac{1}{4}A_{\mu \nu }^{k}A^{\mu \nu k}+\frac{1}{2}m_{A}^{2}A_{\mu
}^{k}A^{\mu k}  \notag \\
&&+\frac{1}{2}(\partial _{a}\pi _{c}^{K})(\partial ^{a}\pi _{c}^{K})-\frac{1%
}{2}m_{\pi }^{2}\pi _{c}^{K}\pi _{c}^{K},  \label{phys}
\end{eqnarray}%
where the corresponding mass terms have also been included
\begin{equation*}
m_{A}^{2}=\frac{36}{N}g^{2}M^{2},\text{ \ }m_{\pi }^{2}=\frac{4}{3N}%
g^{2}M^{2}
\end{equation*}

Finally, let us extract the coupling of the physical pseudoscalar $\pi
_{c}^{K}$ to fermions. Using the axial irreducible component is
\begin{equation*}
T_{abc}^{K}\Big|_{B}=\frac{1}{6}\epsilon _{abcd}B^{dK},
\end{equation*}%
in the relevant tensor part of the gauge-field fermion coupling in (\ref{lfm}%
) one comes to the axial-current coupling
\begin{equation*}
e\mathcal{L}_{B\Psi }=\frac{g}{4}\sqrt{\frac{2}{N}}\,B_{a}^{K}\,J_{5}^{aK},%
\qquad J_{5}^{aK}\equiv \bar{\Psi}\gamma ^{a}\gamma _{5}\lambda _{K}\Psi ,
\end{equation*}%
up to the overall sign fixed by the convention for $\epsilon _{abcd}$ and $%
\gamma _{5}$. Eliminating the auxiliary pseudo-vector $B_{a}^{K}$ through
its algebraic equation of motion gives, to first order in the fermion
current,
\begin{equation}
e\mathcal{L}_{\pi \Psi }=-\frac{\sqrt{3}}{4M}(\partial _{a}\pi _{c}^{K})\bar{%
\Psi}\gamma ^{a}\gamma _{5}\lambda _{K}\Psi .  \label{pipsi}
\end{equation}%
Thus the physical pseudoscalar mode contained in the tensor connection
couples only to the non-conserved part of the hyperflavor axial fermion
current, with a strength suppressed by the TGR scale $M$. The absence of the
gauge coupling constant $g$ in (\ref{pipsi}) is natural: $\pi _{c}^{K}$ is
the canonically normalized longitudinal mode of the heavy tensor connection,
so its derivative coupling is solely controlled by the symmetry-breaking
scale $M$.

\subsection{Hyperflavor-blind vacuum}

In the fixed-norm formulation, the length constraint (\ref{eql}) may be imposed via a Lagrange multiplier. It fixes the norm of the tetrad but leaves its internal orientation
undetermined. A convenient parametrization of the vacuum is \footnotemark[4] 
\begin{equation}
\overline{e}_{\mu }\;=\;v_{\mu }\otimes H,\qquad v_{\mu }=e_{\mu
}^{\,a}\gamma _{a},\qquad H^{\dagger }=H,  \label{eqv}
\end{equation}%
so that the hyperflavor dependence resides entirely in the Hermitian matrix $%
H$. The choice $H\propto \mathbf{1}_{N}$ realizes a hyperflavor-blind vacuum
(HBV) that preserves the residual $SU(N)$ and evidently satisfies (\ref{eql}%
). It is, however, necessary to show that quantum corrections select this
orientation as the true minimum.

The one-loop CW potential for the vector fields in a background $\overline{e}
$ is
\begin{equation*}
U_{V}(e)\;=\;\frac{3}{64\pi ^{2}}\sum_{n}m_{n}^{4}(e)\left( \ln \frac{%
m_{n}^{2}(e)}{\mu ^{2}}-\frac{5}{6}\right) ,
\end{equation*}%
where $m_{n}^{2}$ are eigenvalues of the vector mass matrix $M_{V}^{2}(e)$
and $\mu $ is the renormalization scale. For $SU(N)$ generators $%
G_{V}=\lambda _{k}$ one finds
\begin{eqnarray*}
(M_{V}^{2})_{kl}(e)\; &=&-g^{2}\mathrm{Tr}\!\left( [\lambda _{k},\overline{e}%
_{\mu }][\lambda _{l},\overline{e}^{\mu }]\right) \\
&=&-4g^{2}M^{2}\,\mathrm{Tr}\!\left( [\!\lambda _{k},H][\!\lambda
_{l},H]\right)
\end{eqnarray*}%
where we used
\begin{equation*}
\lbrack \lambda _{k},\overline{e}_{\mu }]\;=\;v_{\mu }\otimes \lbrack
\,\lambda _{k},H\,]\,.
\end{equation*}%
If $H$ is proportional to the identity, all commutators vanish, the vector
mass eigenvalues are zero, and the vector contribution to the CW potential
vanishes after the standard subtraction. For a non-singlet orientation of $H$%
, some vector fields acquire nonzero masses and generate a nontrivial
positive contribution to the curvature of the effective potential around the
hyperflavor-blind point. Thus the vector sector selects the
hyperflavor-blind direction as the preferred vacuum orientation.

The axial and tensor contributions to the CW potential, $U_{A}(e)$ and $%
U_{T}(e)$, do not change the fact that the hyperflavor-blind direction is
selected by the compact vector sector. Near the HBV their leading masses are
set by the scale $gM$ and are largely controlled by the Dirac structure of
the corresponding generators, thus contributing only an additive constant at
$H\propto \mathbf{1}_{N}$. Nevertheless, their dependence on tetrad
fluctuations gives a nonzero curvature of the effective potential in the
hyperflavor-adjoint directions, and thereby generates masses for the
hyperflavor-carrying tetrad modes.

In the post-soldering phase the lower and upper tetrad-type fields are no
longer independent premetric variables. The flavored tetrads, $e_{\mu
}^{\ast ak}$ and $E_{a}^{\ast \mu k}$, as well as the two premetrics $g_{\mu
\nu }$ and $G^{\mu \nu }$, are related as Lorentz-conjugated descriptions of
the same metric geometry. Thus the independent hyperflavor-adjoint tetrad
fluctuations reduce to $16(N^{2}-1)$ components. At the HBV these decompose,
in local Lorentz indices, into antisymmetric and symmetric parts%
\begin{equation}
e_{\mu }^{\ast ak}=\frac{1}{2}e_{\mu b}\left( e^{\ast \lbrack ab]k}+e^{\ast
\{ab\}k}\right)  \label{1/2}
\end{equation}%
where $e_{\mu b}$ represents the corresponding neutral tetrad on the branch (%
\ref{eqv}).

The antisymmetric $6(N^{2}-1)$ modes in (\ref{1/2}), being
Lorentz-rotation-type tetrad components, are absorbed by the tensor gauge
fields $T_{\mu }^{[ab]K}$. Among the symmetric $10(N^{2}-1)$ modes, the $%
(N^{2}-1)$ adjoint traces $e_{\mu }^{\ast \mu k}$ become longitudinal for
the axial vectors $A_{\mu }^{k}$. This leaves $9(N^{2}-1)$ independent
symmetric traceless modes in the $SU(N)$ adjoint, apart from the neutral
tetrad $e_{\mu }^{a}$ in the singlet sector. Its $6$ antisymmetric
components analogously provide masses for the $SL(2,C)$ spin-connections,
while its $10$ symmetric components contain the ordinary metric
perturbation. The role of the TGR combination in (\ref{tr}) is precisely to
give these symmetric neutral components the standard massless spin-2 kinetic
structure, rather than a generic ghostful tetrad kinetic term. Consequently,
one obtains the Fierz-Pauli kinetic operator (\ref{fp}) for the metric
variations $h_{\mu \nu }$. Four components are gauge redundancies of
diffeomorphism invariance, while the components $h_{00}$ and $h_{0i}$ enter
without independent time-derivative kinetic terms and act as constraint
variables. Their equations remove the non-propagating scalar and vector
parts of $h_{\mu \nu }$, so the neutral TGR tetrad sector propagates only
the two transverse-traceless graviton polarizations.

Near the HBV, the axial and tensor loops generate a positive-definite CW
mass matrix along the hyperflavor-adjoint symmetric directions.
Schematically,
\begin{equation*}
M_{A,T}^{2}(e)\;\propto \;g^{2}\,\mathrm{Tr}\!\left( i[G_{A,T},e_{\mu
}]\;i[G_{A,T},e^{\mu }]\right),\qquad G_{A}=\gamma _{5}\lambda
^{k},\;\;G_{T}=\gamma _{ab}\lambda ^{K}
\end{equation*}%
and at quadratic order this gives
\begin{equation*}
\Delta \mathcal{L}_{\mathrm{CW}}^{(2)}\;=\;-\frac{1}{2}e_{\mu }^{\ast ak}(%
\mathcal{M}^{2})_{kl}e_{a}^{\ast \mu l},\qquad (\mathcal{M}^{2})_{kl}\sim
\frac{g^{4}}{16\pi ^{2}}\,M^{2}
\end{equation*}%
for the $9(N^{2}-1)$ flavored symmetric modes. This CW lifting applies only
to the flavored tetrad components $e_{\mu }^{\ast ak}$, while the neutral
gravitational tetrad remains massless. Its masslessness is protected by the
residual spacetime gauge symmetries: the TGR sector gives the usual
Fierz-Pauli kinetic operator for the metric perturbation, but no graviton
mass term. Thus, as required, the gravitational sector is described entirely
by the flavor-singlet tetrad (\ref{or6}), while the flavored tetrad
components are either absorbed by the tensor and axial-vector gauge fields
or acquire large CW masses and thus decouple.

Summarizing this stage: (i) the dynamical tetrad selects the
hyperflavor-blind branch of the theory, thereby breaking the noncompact
directions of $SL(2N,C)$ while preserving the compact hyperflavor group $%
SU(N)$; (ii) on this branch, the tetrad-condensation mechanism filters the
original unified multiplet: the compact vector gauge fields remain massless,
whereas the noncompact axial-vector and tensor directions, together with the
non-singlet tetrad components, decouple from the low-energy sector; (iii) in
general, the theory may contain two independent dimensional scales, $M_{1}$
and $M_{2}$, associated with the independent linear and quadratic curvature
structures. At the fundamental level these scales are arbitrary parameters,
although one or both of them may be related to the Planck scale; and (iv)
this arbitrariness naturally motivates an emergent-gravity scenario, in
which the low-energy Einstein--Cartan operator (\ref{127}) is not
necessarily fundamental, but can be generated radiatively. This possibility
will be discussed in detail in the next section.

\section{Emerging gravity}

\subsection{Preliminaries}

We now turn to the emergence of the EC term from radiative corrections. The
hyperunified $SL(2N,C)$ gauge theory described above is formulated in a
Yang--Mills-like fashion, with all field strengths entering quadratically
and tetrads treated as constrained dynamical fields. The nonlinear constraint (\ref{eql}) fixes the soldering norm, while the dynamically selected hyperflavor-blind branch (\ref{e}) corresponds to the neutral tetrad component (\ref{or6}) and realizes the spontaneous breaking $SL(2N,C)\to SL(2,C)\times SU(N)$. After these steps the low-energy gauge
multiplet consists of the $SU(N)$ vectors and the flavor-singlet tensor
connection $T_{\mu }^{[ab]}$.

At this stage the Lagrangian need not contain a tree-level EC term linear in
the tensor curvature $R[T]$, even though such a term is allowed by symmetry.
Instead, we will show that this term is naturally generated at one loop by
fermion bubbles that involve both the tensor connection and the tetrad. Two
distinct but complementary routes may be considered, depending on the masses
of the fermions contributing to the loops. In a \textquotedblleft
threshold\textquotedblright\ scenario, heavy vectorlike fermions of mass $%
m_{\psi }$ induce a finite, regulator-independent EC term proportional to $%
N_{f}m_{\psi }^{2}$ (for $N_{f}$ species), reminiscent of Sakharov's induced
gravity \cite{Sakharov:1967} and related effective field theory analyses
\cite{Adler:1982ri, Zee:1981, Don}, although such a mechanism has not yet been 
applied to gravity-unified theories. In a \textquotedblleft universal
cutoff\textquotedblright\ scenario, if gravity or some UV completion
provides a physical cutoff $\Lambda \sim M_{P}$ for the relevant loop
integrals, even light fermions induce an EC term of order $N_{f}\Lambda ^{2}$%
; this is regulator-dependent but fixes the correct overall scale. In both
pictures the EC term becomes radiative in origin, expressed in terms of the
underlying $SL(2N,C)$ parameters while maintaining the unified structure of
the quadratic sector.

\subsection{Radiative corrections}

After tetrad condensation, the quadratic curvature sector is governed by the
ghost-free Lagrangian $\mathcal{L}_{G}^{(2)}$ (\ref{re}) embedded into the $%
SL(2N,C)$ framework. A separate linear curvature term of the EC form for the
surviving flavor-singlet tensor field submultiplet
\begin{equation}
e\mathcal{L}_{\mathrm{EC}}=\frac{1}{2\kappa }\,\,e_{[a}{}^{\mu
}e_{b]}{}^{\nu }R_{\mu \nu }^{ab}{}[T],  \label{eq:EC}
\end{equation}%
is allowed by symmetry but need not appear at tree level. We show that this
structure---with $R_{\mu \nu }^{ab}{}[T]$ denoting the curvature of the
flavor-singlet tensor connection $T_{\mu }^{[ab]}$---is generated
radiatively by fermion loops in which both the tetrad and the tensor field
couple to matter.

The relevant interaction vertices follow from the fermion kinetic term and
the $SL(2N,C)$ covariant derivative (\ref{lfm}) when specialized to the
neutral tensor field. The basic insertions in the fermion loops are
\begin{align*}
V_{e}& :\ \,ie^{\mu }{}_{a}\,\bar{\Psi}\,\gamma ^{a}\overleftrightarrow{%
\partial _{\mu }}\Psi , \\
V_{T}& :\ \frac{i\,g}{4}\,e^{\mu }{}_{c}\,\bar{\Psi}\,\gamma ^{c}\gamma
_{ab}\,\Psi \,T_{\mu }^{[ab]}\,
\end{align*}%
Correspondingly, the vertex $V_{e}$ brings an external momentum, while the
vertex $V_{T}$ insertion involves both an external tetrad leg and an
external $T$ leg with the coupling constant $g$.

The minimal one-loop origin of the operator $e\wedge e\wedge R[T]$ is then a
pair of fermion-bubble diagrams, illustrated in Fig.~\ref{fig:bubbles}. The first contains one pure tetrad
insertion $V_{e}$ and one mixed tensor--tetrad insertion $V_{T}-V_{e}$,
producing a three-point function corresponding to $e\wedge e\wedge dT$. The
second contains two mixed $V_{T}-V_{e}$ insertions, producing $e\wedge
e\wedge (g\,T\wedge T)$. Taken together they reconstruct the full $e\wedge
e\wedge R[T]$ structure with $R[T]=dT+gT\wedge T$.

\begin{figure}[!t]
\centering
\begin{tikzpicture}[baseline=(c.base),scale=0.85,transform shape]
            \begin{feynman}
                \vertex (c) at (0,0);
                \vertex (d) at (1.6,0);
                \diagram*{
                    (c) -- [fermion, half left] (d)
                    -- [fermion, half left] (c)
                };
                \vertex[left=0.7cm of c] (Te) {};
                \vertex[below=0.6cm of c] (Ee) {};
                \diagram*{
                    (Te) -- [boson] (c),
                    (Ee) -- [scalar, dashed] (c)
                };
                \node at ($(Te)!0.25!(c)+(0,0.2)$) {\scriptsize $T$};
                \node at ($(Ee)!0.3!(c)+(-0.18,0.05)$) {\scriptsize $e$};
                \vertex[right=0.7cm of d] (Eer) {};
                \diagram*{
                    (Eer) -- [scalar, dashed] (d)
                };
                \node at ($(Eer)!0.25!(d)+(0,0.2)$) {\scriptsize $e$};
            \end{feynman}
        \end{tikzpicture}
\hspace{0.7cm}
\begin{tikzpicture}[baseline=(c2.base),scale=0.85,transform shape]
            \begin{feynman}
                \vertex (c2) at (0,0);
                \vertex (d2) at (1.6,0);
                \diagram*{
                    (c2) -- [fermion, half left] (d2)
                    -- [fermion, half left] (c2)
                };
                \vertex[left=0.7cm of c2] (T1) {};
                \vertex[below=0.6cm of c2] (E1) {};
                \vertex[right=0.7cm of d2] (T2) {};
                \vertex[below=0.6cm of d2] (E2) {};
                \diagram*{
                    (T1) -- [boson] (c2),
                    (E1) -- [scalar, dashed] (c2),
                    (T2) -- [boson] (d2),
                    (E2) -- [scalar, dashed] (d2)
                };
                \node at ($(T1)!0.25!(c2)+(0,0.2)$) {\scriptsize $T$};
                \node at ($(E1)!0.3!(c2)+(-0.18,0.05)$) {\scriptsize $e$};
                \node at ($(T2)!0.25!(d2)+(0,0.2)$) {\scriptsize $T$};
                \node at ($(E2)!0.3!(d2)+(0.18,0.05)$) {\scriptsize $e$};
            \end{feynman}
        \end{tikzpicture}
\caption{Fermion-bubble diagrams generating the Einstein--Cartan term. Left:
one pure tetrad insertion $V_{e}$ and one mixed $V_{T}$--$V_{e}$ insertion
generate $e\wedge e\wedge dT$. Right: two mixed $V_{T}$--$V_{e}$ insertions
generate $e\wedge e\wedge (gT\wedge T)$. Their sum reconstructs $e\wedge
e\wedge R[T]$.}
\label{fig:bubbles}
\end{figure}
The coefficient of \eqref{eq:EC} depends on the fermion content and on the
UV completion. Two complementary scenarios are illustrative. In a
\textquotedblleft threshold\textquotedblright\ scenario, vectorlike fermions
with masses $m_{\psi }\sim M_{P}$ generate a regulator-independent
contribution
\begin{equation*}
e\mathcal{L}_{\mathrm{EC}}^{(t)}\sim \frac{N_{f}}{16\pi ^{2}}\,m_{\psi
}^{2}\,e\wedge e\wedge R[T],
\end{equation*}%
so that $M_{P}^{2}\sim (N_{f}/16\pi ^{2})m_{\psi }^{2}$. In a
\textquotedblleft universal cutoff\textquotedblright\ scenario, quantum
gravity itself provides a physical cutoff $\Lambda \sim M_{P}$, and even
light (or massless) fermions induce
\begin{equation*}
e\mathcal{L}_{\mathrm{EC}}^{(c)}\sim \frac{N_{f}}{16\pi ^{2}}\,\Lambda
^{2}\,e\wedge e\wedge R[T],
\end{equation*}%
with the Planck scale determined by the UV boundary condition set by the UV
completion. In both cases the emergent EC term is expressed in terms of the
parameters of the underlying $SL(2N,C)$ gauge theory and its fermion sector.
The quadratic curvature terms are already present via the GFC type
Lagrangian, so the full low-energy gravitational sector is an
EC-plus-curvature-squared theory with the correct sign and no ghosts.

Analogously, the TGR term is generated instead by torsion-square-type
fermion loop structures with insertions
\begin{eqnarray*}
V_{e}-V_{e}\quad &\longrightarrow &\quad (\partial e)^{2}, \\
V_{e}-V_{T}\quad &\longrightarrow &\quad (\partial e)(Te) \\
V_{T}-V_{T}\quad &\longrightarrow &\quad (Te)^{2}
\end{eqnarray*}%
Together these reconstruct the tensor-connection-covariant part of the TGR
torsion%
\begin{equation*}
C_{\mu \nu }^{a}=C_{\mu \nu }^{a}(e)+C_{\mu \nu }^{a}(T)
\end{equation*}%
and hence the corresponding torsion scalar operator $T(C)$ in (\ref{tr}, \ref%
{ltr1}). On the strict hyperflavor-blind branch, the direct axial fermion
vertex vanishes. Therefore the fermion determinant on that branch need not
generate the axial-vector lifting part $C_{\mu \nu }^{a}(A)$ of the TGR
torsion.

\subsection{Concluding remarks}

These considerations show that the linear curvature term allowed by $%
SL(2N,C) $ is generically a sum of a bare piece and an induced one-loop
contribution. In a threshold scenario, superheavy vectorlike fermions near
the Planck scale generate a finite EC term which may dominate over any
tree-level coefficient if the latter is tied to lower mass scales. The
Planck mass may then be viewed as largely radiative in origin, controlled by
the multiplicity and masses of heavy fermion states.

In the universal-cutoff picture, quantum gravity or some UV completion
supplies a physical scale $\Lambda \sim M_{P}$ that regularizes the relevant
loop integrals. The induced EC term from light or massless fermions then
takes the form $\Delta M^{2}\propto N_{f}\Lambda ^{2}$, so the observed
Planck mass squared $M_{P}^{2}$ can be viewed as the sum $M^{2}+\Delta M^{2}$%
, where $M^{2}$ is a bare parameter. Assuming $M^{2}$ is tied to much lower
scales (e.g.\ the $SL(2N,C)$ breaking scale or the weak scale) makes the
induced contribution dominant.

One may further entertain the idea that the $SL(2N,C)$ theory (including its
SM sector) is classically-scale invariant, with all large scales generated
radiatively or via the UV completion. In such a setting the EC term is
absent at tree level, so that the Planck scale and the effective EC coupling
necessarily arise solely from such radiative corrections. This does not, by
itself, solve the hierarchy problem, but it clarifies how a large
gravitational scale may arise without introducing Planck-scale VEVs inside
the $SL(2N,C)$ theory.

Finally, let us comment on the radiative stability of the ghost-free TGR and
GFC frameworks used above. These structures should be regarded as the
unitary kinetic projectors of the two dynamical sectors, not as accidental
tunings among arbitrary quadratic invariants. The TGR torsion term selects
the tetrad kinetic structure that propagates only the two massless spin-2
graviton polarizations, eliminating the would-be ghostlike scalar, vector,
and antisymmetric tetrad components. The curvature-square GFC plays the
analogous role for the spin connection: its special structure projects out
the unwanted ghostlike connection components. In a unitary quantum theory,
radiative corrections may renormalize the overall coefficients of these
admissible projectors, and may also generate higher-dimensional operators
suppressed by the ultraviolet scale, but they cannot reintroduce uncancelled
negative-norm poles into the physical spectrum. Any same-order deformation
away from the TGR or GFC surfaces that would produce such poles must
therefore be absent, cancelled by the full set of quantum contributions, or
fixed by the unitarity-preserving renormalization conditions. This is the
sense in which the TGR and GFC frameworks are radiatively stable within the
unitary effective theory.

\section{SL(2N,C) breaking scenario}

\subsection{Post-soldering phase}

In general, the logic of the broken phase should not be understood as a
truncation of the original $SL(2N,C)$ theory to an ordinary $SL(2,C)$
theory. Rather, we keep the full $SL(2N,C)$-covariant structure, while the
soldering branch selects the hyperflavor-blind direction (\ref{or6}). Thus
the physical spacetime metric is determined by the neutral tetrad $e_{\mu
}^{a}$, but the covariant derivatives acting on all multiplets remain the
full $SL(2N,C)$ covariant derivatives. In particular, the covariant
derivative for a tetrad in (\ref{Cmunu}) is written with the complete $%
SL(2N,C)$ connection. This is the standard Higgs-phase logic: the vacuum
chooses a direction in group space, but the broken gauge fields still appear
through the covariant derivative of that vacuum. Consequently, although the
tetrad is neutral in hyperflavor space, the full connection still probes the
broken directions through commutators with the neutral tetrad.

This same principle is used in all sectors of the theory. First, in the
tetrad sector, the TGR torsion scalar is interpreted as the ghost-free
kinetic projector for the dynamical tetrad. On the hyperflavor-blind branch,
it gives the healthy massless spin-2 kinetic structure for the neutral
tetrad fluctuations and removes the would-be ghostlike scalar, vector, and
antisymmetric tetrad components. The essential requirement is therefore the
TGR no-ghost structure for the physical neutral tetrad sector. The full $%
SL(2N,C)$-covariant derivative may also contain lifting terms for the broken
connection components, but the absence of a particular radiatively induced
axial-vector contribution does not by itself reintroduce ghostlike tetrad
modes.

Second, the spin-connection sector is treated analogously. The full $%
SL(2N,C) $ gauge multiplet is retained, but its Yang--Mills-type kinetic
term must be chosen in the generalized GFC combination. This special
curvature-square structure is the no-ghost kinetic projector for the spin
connections, just as TGR is the no-ghost kinetic projector for the tetrad.
Thus the $SL(2N,C)$ gauge fields are not arbitrary higher-derivative
connection modes; their healthy propagation is defined by the GFC-type
structure.

Third, the same broken-phase logic applies to induced gravitational terms.
The radiative corrections are to be computed from the full matter coupling,
rather than from an artificially neutral-truncated set of interaction
vertices. After the effective action is obtained, one evaluates it on the
hyperflavor-blind soldering vacuum. In this way the induced EC term, and
likewise the induced TGR-type tetrad kinetic term, should be understood as
broken-phase remnants of the full $SL(2N,C)$ covariant matter theory. The EC
term is generated by the familiar fermion-loop mechanism involving tetrad
and spin-connection insertions, while the TGR term corresponds to the
induced ghost-free kinetic operator for the neutral dynamical tetrad. The
loop calculation need not generate every possible broken-connection lifting
term in the covariant torsion; what is essential is that the induced tetrad
kinetic sector lies on the TGR no-ghost surface.

In this sense the construction is internally consistent. The
hyperflavor-blind tetrad vacuum provides the ordinary spacetime geometry,
while the full $SL(2N,C)$-covariant derivatives keep track of the broken
gauge directions and their mass generation. The TGR and GFC structures then
supply the two required ghost-free kinetic projectors: TGR for the dynamical
tetrad and GFC for the spin connection. Radiative fermion loops can induce
the low-energy EC and TGR gravitational operators in this broken phase,
while the no-ghost structure fixes which quadratic combinations are
physically admissible. Thus one obtains a gravity-inspired hyperunified $%
SL(2N,C)$ theory without reducing the parent gauge-covariant framework to an
ordinary neutral $SL(2,C)$ theory.

\subsection{Interplay of two gravities}

In the hyperflavor-blind broken phase, the tetrad selects the neutral
direction (\ref{or6}) inside the $SL(2N,C)$ group space, but the theory is
not reduced to an ordinary $SL(2,C)$ theory. The covariant derivatives
acting on all multiplets still contain the full $SL(2N,C)$ connection $%
I_{\mu }$. Thus EC gravity and TGR gravity in this framework use the same
neutral tetrad and the same full spin-connection multiplet, but they
organize the gravitational dynamics differently.

The EC type term is curvature-based,
\begin{equation*}
e\mathcal{L}_{\mathrm{EC}}\sim M_{\mathrm{EC}}^{2}\,\mathrm{Tr}(e^{\mu
}e^{\nu }I_{\mu \nu })
\end{equation*}%
where $I_{\mu \nu }$ is the curvature of the full $SL(2N,C)$ connection,
projected onto the neutral Lorentz-tetrad direction. Meanwhile, the TGR type
term is torsion-based,
\begin{equation*}
e\mathcal{L}_{\mathrm{T}}\sim M_{\mathrm{T}}^{2}\,\mathrm{Tr}\mathbb{T}(C)
\end{equation*}%
as presented above in Section 3.4. Therefore, EC measures the curvature of
the total connection, while TGR measures the covariant non-integrability of
the neutral tetrad under the same total connection. In the limit in which
the extra broken connection components are heavy or decouple, both
descriptions reduce to the same low-energy neutral metric gravity: the same
massless spin-2 graviton, the same Newtonian limit, light bending, redshift,
and ordinary gravitational-wave propagation. Away from this low-energy
neutral limit, they differ through their coupling to the full
spin-connection multiplet.

The main difference relevant here appears in the mass terms of the gauge
submultiplets, when the generalized GFC term supplies the quadratic kinetic
operator for the full $SL(2N,C)$ spin-connection multiplet. Though both
gravity operators, $e\mathcal{L}_{\mathrm{EC}}$ and $e\mathcal{L}_{\mathrm{T}%
}$, leave the vector submultiplets massless, they generate different mass
patterns for the axial-vector and tensor submultiplets. The EC term, on the
hyperflavor-blind tetrad branch, $e_{\mu }=e_{\mu }^{a}\gamma _{a}\lambda
^{0}$, projects the full connection curvature onto the Lorentz-tensor part.
Its quadratic expansion gives an algebraic mass term of the order $M_{%
\mathrm{EC}}^{2}$ for the tensor spin-connection submultiplet $T_{\mu
}^{[ab]K}$, while the axial-vector submultiplet $A_{\mu }^{k}$ does not
enter the neutral EC projection at quadratic order, and thereby is not
lifted by the EC term itself. By contrast, the TGR term contains the full
covariant torsion of the neutral tetrad. Since the broken tensor and
axial-vector generators do not commute with the neutral tetrad term, the TGR
torsion contains both tensor and axial-vector lifting pieces proportional to
$M_{\mathrm{T}}^{2}$.

\subsection{Achieving hyperunification}

The existence of an $SL(2N,C)$ gauge structure by itself does not
automatically imply a full dynamical unification of gravity with the other
elementary interactions. The symmetry organizes the Lorentz, tensor,
axial-vector, and internal gauge fields into one $SL(2N,C)$ connection $%
I_{\mu }$, whose universal Yang--Mills-type kinetic term must be chosen in
the generalized GFC form. However, once tetrad multiplets are present, the
same general $SL(2N,C)$ covariant framework also permits independent
gravitational invariants built from the tetrads and the full connection.
Thus, besides the GFC gauge kinetic terms $\mathcal{L}_{\mathrm{GFC}}$
presented in terms of the propagating fields in (\ref{phys}), one may also
write independent EC type and TGR type terms, $\mathcal{L}_{\mathrm{EC}}$
and $\mathcal{L}_{\mathrm{T}}$, with their own dimensionful coefficients, $%
M_{\mathrm{EC}}^{2}$ and $M_{\mathrm{T}}^{2}$. In general, these terms are
allowed by the starting $SL(2N,C)$ structure already in the premetric
invariant phase, and are then reduced on the hyperflavor-neutral tetrad
branch discussed above. Therefore, if $M_{\mathrm{EC}}^{2}$ and $M_{\mathrm{T%
}}^{2}$ are arbitrary bare parameters, the theory is not yet a true
dynamical unification of gravity with the other forces. It is a unification
at the level of covariant derivatives: the spin-connection and internal
symmetry objects are embedded into the same $SL(2N,C)$ gauge multiplet. But
the gravitational strengths are still inserted independently through the
separate mass-squared coefficients $M_{\mathrm{EC}}^{2}$ and $M_{\mathrm{T}%
}^{2}$. In that case the Planck scale is not explained by the unified gauge
structure; it is put in by hand.

A genuine hyperunification interpretation requires an additional dynamical
assumption. The fundamental bare EC and TGR terms should be absent,
suppressed, or insignificant compared with the induced terms generated by
the heavy matter sector. Then the dimensionful gravitational coefficients
are not independent input parameters. Instead, they arise radiatively from
fermion loops involving the same $SL(2N,C)$ covariant matter couplings (\ref%
{lfm}) that also define the unified gauge interaction. In this case the
induced EC and TGR coefficients inherit the same universal gauge coupling $g$
that appears in the covariant derivative, in the $SL(2N,C)$ curvature, and
in the generalized GFC term. Schematically, in the chosen normalization,
\begin{equation*}
M_{\mathrm{EC}}^{2},\;M_{\mathrm{T}}^{2}\sim \frac{N_{f}}{16\pi ^{2}}\,g^{2}%
\boldsymbol{M}^{2},
\end{equation*}%
up to numerical coefficients, where $\boldsymbol{M}$ denotes the heavy
fermion threshold or cutoff scale. The bare coefficients, $M_{\mathrm{EC,0}%
}^{2}$ and $M_{\mathrm{T,0}}^{2}$, may then be taken much smaller than the
induced threshold contributions, so that the observed gravitational scales
are generated by the same universal gauge-coupled matter sector rather than
imposed independently. There is, however, an important distinction between
the bare $M_{\mathrm{EC,0}}^{2}$ and $M_{\mathrm{T,0}}^{2}$ coefficients. $%
M_{\mathrm{EC,0}}^{2}$ merely shifts the effective EC scale and can be
absorbed into the definition of the total induced gravitational constant. By
contrast, $M_{\mathrm{T,0}}^{2}$ not only contributes to this constant, but
also supplies a mass lifting term for the axial-vector submultiplet $A_{\mu
}^{k}$, which would remain massless in the absence of the corresponding TGR
lifting.

This clarifies the meaning of hyperunification. The universal GFC term
supplies the no-ghost kinetic structure for the full $SL(2N,C)$ gauge
multiplet and contains the same gauge coupling $g$ that controls the matter
interactions. The EC and TGR terms are allowed by the same symmetry, but if
their coefficients are arbitrary bare couplings, they weaken the claim of
dynamical unification. The stronger hyperunification scenario is obtained
only when these gravitational terms are radiatively induced from the same $%
SL(2N,C)$ covariant fermion sector. Then the Planck scale gravitational
dynamics originate from the same high-energy gauge-coupled matter dynamics.

The ultraviolet interpretation should therefore be that the fundamental
theory is a fully $SL(2N,C)$-invariant premetric theory with massless gauge
fields and no prior ordinary spacetime metric. The tetrad multiplets define
the premetric soldering sector, while the full connection $I_{\mu }$ is
governed by the generalized GFC kinetic term. In the infrared, the theory
selects the hyperflavor-blind neutral tetrad branch and breaks to the
ordinary spacetime phase with the $SL(2,C)\times SU(N)$ symmetry as the
surviving gauge structure. The $SU(N)$ vector fields $V_{\mu }^{k}$ remain
massless and the axial-vector multiplet $A_{\mu }^{k}$ becomes massive at
the generic TGR scale through the covariant-torsion lifting term. Meanwhile,
the tensor connection sector $T_{\mu }^{[ab]K}$ produces the massive
pseudoscalar multiplet $\pi ^{K}$ of order the Planck scale. With the TGR
and GFC chosen as the no-ghost kinetic projectors, and with positive induced
mass-squared coefficients, the infrared broken-phase theory is ghost- and
tachyon-free, as summarized in (\ref{phys}). The ordinary gravitational
scale is not a fundamental independent constant, but an induced threshold
effect of the same $SL(2N,C)$ covariant matter sector. This is the sense in
which the construction becomes a genuine gravity-inspired hyperunification
rather than merely a representation-level embedding of gravity and internal
gauge fields into a larger group. The most important point is that genuine
gravitational unification emerges only in the broken post-soldering phase,
in contrast to ordinary grand unification of internal forces, which has both
an exact high-symmetry limit and a broken low-energy realization.

\section{From gauge to matter sector: true HUT carriers}

\subsection{Are quarks and leptons truly elementary?}

So far we have concentrated on the gauge sector. We now address the matter
sector and its relation to hyperunification. After tetrad condensation the
effective local symmetry is $SL(2,C)\times SU(N)$, while the full $SL(2N,C)$
algebra continues to organize multiplets of fields. On the one hand this
yields $SL(2,C)$ gauge gravity; on the other, it provides $SU(N)$ as the
candidate grand-unified group. However, because $SL(2N,C)$ multiplets carry
both spinor and internal indices, such GUT constructions do not accommodate
ordinary quarks and leptons comfortably. The spin-$1/2$ fundamental
representation of $SL(2N,C)$ has quantum numbers that do not match the
observed pattern of Standard Model charges, while higher representations
generally mix fields of different spins inside a single multiplet and
therefore are not suitable for describing purely fermionic states\footnote{%
Thus the matter-sector question is not merely how to embed known quarks and
leptons into $SL(2N,C)$, but which truly elementary spinor fields can
consistently carry the full hyperunified symmetry.}.

This can be seen already in the simplest case of $SU(5)$, which might be
expected to arise from an $SL(10,C)$ hyperunification scheme. Some
low-dimensional chiral multiplets, say left-handed ones transforming under $\Omega_L$ in (\ref{lr}), may then be written in $SU(5)\times SL(2,C)$ components as
\begin{equation*}
\Psi _{L}^{ia}\text{ },\text{\ \ \ }10=(\overline{5},2)
\end{equation*}%
and
\begin{equation}
\Psi _{L[ai,\text{ }jb]}=\Psi _{L[ij]\{ab\}}+\Psi _{L\{ij\}[ab]},\text{ \ }%
45=(10,\text{ }3)+(15,\text{ }1)\,.  \label{sl2}
\end{equation}%
Here $i,j=1,\dots ,5$ label $SU(5)$ indices and $a,b=1,2$ label $SL(2,C)$
spinor indices. In these decompositions, antisymmetry in the combined $(ia)$
indices implies antisymmetry in $i,j$ and symmetry in $a,b$, and conversely.
One finds that the $\overline{5}$ representation appears naturally, but the
usual fermionic decuplet of $SU(5)$ does not emerge from a purely
antisymmetric $SL(10,C)$ representation: the tensor (\ref{sl2}) instead
corresponds to vector and scalar multiplets rather than to a fermionic
decuplet. As a result, the standard $SU(5)$ GUT \cite{gg}, together with
many of its supersymmetric \cite{su} or string-inspired extensions \cite{98,
99}, cannot be straightforwardly realized in this hyperunified setting.

More generally, GUTs in which quarks and leptons are placed in purely
antisymmetric representations (such as $SU(11)$ models \cite{ge}) are
problematic: the spin content of the resulting states does not match that of
observed fermions once the requirement of an underlying $SL(2N,C)$ structure
is imposed. To maintain spin-$1/2$ assignments for ordinary quarks and
leptons within an $SL(2N,C)$ framework one would need more complicated
multiplets involving both upper and lower indices, which tend to be
extremely large and introduce many exotic states.

This strongly points toward a composite interpretation: quarks and leptons
should not be identified with fundamental $SL(2N,C)$ spinors, but may
instead emerge as bound states of more fundamental chiral preons. For such
preons the $SL(2N,C)$ representation theory becomes much more economical,
with preons transforming in fundamental spinor representations and
composites living in higher-rank tensors. In the remainder of this section
we follow this line of thought and argue that an $SL(16,C)$ hyperunified
model with $SU(8)$ hyperflavor provides a natural home for three generations
of composite quarks and leptons \cite{jpl}.

\subsection{Preons: metaflavors and metacolors}

We adopt the preon framework developed in \cite{ch}. At very short
distances, potentially near the Planck scale, we postulate the existence of $%
2N$ massless preons, $N$ left-handed and $N$ right-handed, described by
independent Weyl spinors $P_{iaL}$ and $Q_{iaR}$ with $i=1,\dots ,N$ and $%
a=1,2$. They carry a common local metaflavor symmetry
\begin{equation*}
G_{MF}=SL(2N,C)\,,
\end{equation*}%
acting in the fundamental representation on both $P_{iaL}$ and $Q_{iaR}$.
This is the same hyperunified group that organized the gauge and tetrad
sectors above, now acting on the truly fundamental matter fields. At the
fundamental level we take this metaflavor symmetry to be vectorlike, so that
$SL(2N,C)$ acts identically on left- and right-handed preons; left--right
symmetry is therefore exact in the preon sector. At large distances,
however, the effective symmetry realized on composite states may be reduced
to chiral subgroups.

In addition to metaflavor, preons transform under a chiral metacolor gauge
group,
\begin{equation*}
G_{MC}=SO(n)^{L}\times SO(n)^{R}\,,
\end{equation*}%
in whose vector representation they reside. We denote the corresponding
indices by $\alpha $ and $\alpha ^{\prime }$ for the left and right
metacolor factors, respectively, and write $P_{iaL}^{\alpha }$ and $%
Q_{iaR}^{\alpha ^{\prime }}$ with $\alpha ,\alpha ^{\prime }=1,\dots ,n$.
The chiral metacolor interactions confine preons into bound states, among
which quarks, leptons, and possible additional composites may appear with a
characteristic confinement radius
\begin{equation*}
R_{MC}\sim 1/\Lambda _{MC}\,,
\end{equation*}%
where $\Lambda _{MC}$ is the scale at which metacolor becomes strong. By
assumption, left and right metacolor sectors have comparable confinement
scales due to the underlying L--R symmetry.

The choice of orthogonal metacolor has several advantages. Vector
representations of $SO(n)$ are free of gauge anomalies, consistent with the
preon assignment, and orthogonal groups allow composites in a variety of
representations of the metaflavor group $SL(2N,C)$, including tensors with
both upper and lower indices. Under the combined symmetry $SL(2N,C)\times
SO(n)^{L}\times SO(n)^{R}$ the preons transform as
\begin{equation*}
P_{iaL}^{\alpha }[N,(1/2,0);(n,1)]\,,\qquad Q_{iaR}^{\alpha ^{\prime
}}[N,(0,1/2);(1,n)]\,,
\end{equation*}%
where we have indicated the transformation properties under metaflavor,
Lorentz, and metacolor factors.

In the limit where metaflavor gauge interactions are formally turned off,
the preon system exhibits an additional global chiral symmetry,
\begin{equation}
K(N)=SU(N)_{L}\times SU(N)_{R}\,,  \label{ch}
\end{equation}
acting independently on left- and right-handed preons. A bilinear condensate
$\langle P_L Q_R\rangle$ that would generate large Dirac masses of order $%
\Lambda_{MC}$ for composites is forbidden by this chiral symmetry. This is a
necessary (though not sufficient) condition for the existence of massless
composite fermions. In fact, massless composites are expected only when the
preon chiral symmetry (\ref{ch}) survives at large distances, and the
corresponding pattern is constrained by the 't Hooft anomaly matching (AM)
condition \cite{t}. Roughly speaking, the $SU(N)^3_{L,R}$ anomalies carried
by preons must match those of the composite fermions in each chiral sector.

An important distinction arises between the vectorlike metaflavor symmetry
and the chiral spectator symmetry (\ref{ch}). For the local $SL(2N,C)$,
gauge anomalies cancel automatically between left- and right-handed preons
and their composites since the theory is vectorlike. For the global chiral
symmetry $SU(N)_{L,R}$, however, anomalies must be matched separately in the
left and right sectors by suitable combinations of preons and massless
composites. In practice one can imagine gauging $SU(N)_{L,R}$ as a spectator
symmetry and introducing additional spectator fermions as in \cite{t}. The
AM condition then constrains the number of preon species $N$ and the
representations in which the massless composite states transform.

\subsection{Composites: $SL(16,C)$ and residual $SL(2,C)\times SU(8)$}

The metaflavor dynamics is governed by the local vectorlike group $SL(2N,C)$
, while the chiral symmetry $SU(N)_L\times SU(N)_R$ remains global. To
analyze anomaly constraints one promotes the chiral symmetry temporarily to
a spectator gauge symmetry with auxiliary gauge fields, allowing one to
compute the $SU(N)_L^3$ and $SU(N)_R^3$ anomalies produced by preons and
composites. In this auxiliary theory anomalies in the $SU(N)_L$ and $SU(N)_R$
sectors are cancelled by suitable spectra of spectator fermions, but the AM
condition enforces equality of the anomalies carried by preons and by
composites in each sector separately.

In the $SL(2N,C)$ metaflavor theory, the cubic $SL(2N,C)^3$ anomalies cancel
between left- and right-handed states for any $N$, since the theory is
vectorlike. The nontrivial constraint comes instead from the chiral symmetry
(\ref{ch}). Let $a(N)$ be the anomaly coefficient for $SU(N)$ in the
fundamental representation (so $a(N)=\pm1$ for left- and right-handed
preons), and let $a(r)$ denote the anomaly coefficient for a given composite
representation $r$. If $i_r$ counts the net number of massless composites in
representation $r$ (positive for left-handed, negative for right-handed
states), the AM condition takes the form
\begin{equation}
n\,a(N)=\sum_{r}i_{r}\,a(r)\, ,  \label{am}
\end{equation}
where $n$ is the number of preons of a given chirality. The coefficients $%
a(r)$ depend on $N$, so (\ref{am}) can in principle fix $N$ once a
particular pattern of composite representations is specified.

To sharpen the constraints we impose three assumptions: (\textbf{i}) all
massless composites contributing to (\ref{am}) are spin-$1/2$; (\textbf{ii})
only the simplest three-preon composites formed by metacolor interactions
are present; (\textbf{iii}) they form a single irreducible representation of
the hyperunified symmetry, which at low energies reduces to a single
irreducible representation of $SL(2,C)\times SU(N)$. Assumption (i) means
that higher-spin components of generic $SL(2N,C)$ multiplets must become
massive and decouple, which in turn implies that the preon metaflavor
symmetry is effectively reduced to $SL(2,C)\times SU(N)$ at large distances,
consistent with the gauge-sector analysis of Secs. 2--4. Assumptions (ii)
and (iii) significantly restrict the possible composite spectra.

Under these assumptions, only three-preon composites built from $P_{L}$ (and
analogously from $Q_{R}$) are considered. The possible $SU(N)$
representations for spin-$1/2$ composites are third-rank tensors of the form
\begin{equation*}
\Psi _{L,R}^{\{ijk\}}\,,\quad \Psi _{\lbrack ijk]L,R}\,,\quad \Psi
_{\{[ij]k\}L,R}\,,\quad \Psi _{\{jk\}L,R}^{i}\,,\quad \Psi _{\lbrack
jk]L,R}^{i}\,,
\end{equation*}%
with symmetrization or antisymmetrization in the flavor indices $%
i,j,k=1,\dots ,N$. Evaluating the anomaly coefficients $a(r)$ for these
representations and imposing (\ref{am}) with $n=3$, $a(N)=1$, and $i_{r}=1$
for a single composite representation $r=r_{0}$ yields
\begin{equation*}
3=a(r_{0})\,.
\end{equation*}%
One finds that an integer solution for $N$ exists only when $r_{0}$ is the
mixed tensor $\Psi _{\,[jk]L}^{i}$ (and its right-handed analogue) and,
moreover, only for a specific value of $N$:
\begin{equation}
3=\frac{N^{2}}{2}-\frac{7N}{2}-1\,,\qquad N=8\,,  \label{8-1}
\end{equation}%
which corresponds to a $216$-dimensional representation for each chirality.
Thus, among all candidate values, only $SU(8)_{L}\times SU(8)_{R}$ can be
realized on massless three-preon composites consistent with anomaly
matching, thereby pointing uniquely, within these assumptions, to $SL(16,C)$
as the underlying metaflavor symmetry\footnote{%
Remarkably, due to the direct screening effect of the orthogonal metacolor,
the AM condition (\ref{8-1}) can be extended to the $SO(n)^{L}\times
SO(n)^{R}$ metacolor symmetry case by supplementing the above three-preon
representation $\Psi _{\,[jk]L}^{i}$ with $p$ composite fundamental
one-preon multiplets $\Psi _{iL}$, appropriately screened by metagluons.
Thus, the new AM condition acquires the form
\begin{equation*}
n=\frac{N^{2}}{2}-\frac{7N}{2}-1+p
\end{equation*}%
whose solutions exist only if $n-p=3$, so the same eightfold chiral symmetry
$SU(8)_{L}\times SU(8)_{R}$ is selected even in the general case \cite{ch}.}.

The unique solution of the AM condition with three-preon composites thus
picks out $N=8$ and identifies
\begin{equation*}
SL(16,C)\longrightarrow SL(2,C)\times SU(8)
\end{equation*}
as the relevant metaflavor pattern: the full $SL(16,C)$ is exact at the
preon level, whereas the residual $SL(2,C)\times SU(8)$ is the symmetry
realized on massless composites at large distances. The corresponding
massless multiplets $\Psi_{\,[jk]aL}^{i}$ and $\Psi_{\,[jk]aR}^{\prime\,i}$
fill the $216_{L,R}$ representations of $SU(8)$, with spinor index $a=1,2$.

Upon decomposing the $SU(8)$ multiplet under $SU(5)\times SU(3)$ one obtains
(suppressing spinor indices)
\begin{equation}
216_{L,R}=[(\overline{5}+10,\text{ }\overline{3}
)+(45,1)+(5,8+1)+(24,3)+(1,3)+(1,\overline{6})]_{L,R}  \label{216}
\end{equation}
where $(\overline{5}+10,\overline{3})_{L}$ corresponds precisely to three
generations of quarks and leptons forming a triplet of a family symmetry $%
SU(3)$ \cite{h}. The same pattern appears for $216_{R}$ due to the
underlying vectorlike $SL(16,C)$ symmetry. Even though preons remain
massless, all composite states listed in (\ref{216}) are in principle
allowed to pair up and acquire heavy masses through Dirac terms.

\subsection{L--R symmetry violation: chiral quark-lepton families}

To obtain a chiral low-energy spectrum containing three light quark-lepton
families, one needs to break the preon-level L-R symmetry effectively, so
that only the left-handed combination $(\overline{5}+10,\overline{3})_{L}$
remains massless, while its right-handed counterpart acquires a large mass.
Following \cite{ch}, we assume that this breaking is triggered by
interactions in the right-handed preon sector, such that the chiral symmetry
realized on right-handed composites at large distances is reduced to $%
[SU(5)\times SU(3)]_{R}$, while the left-handed sector retains the full $%
SU(8)_{L}$.

The corresponding breaking pattern may be caused by a typical L-R symmetric
polynomial potential for the third-rank antisymmetric scalars $\Phi
_{\lbrack pqr]}$ and $\Phi _{\lbrack pqr]}^{\prime }$ composed from $P$ and $%
Q$ preons, respectively. This potential, after breaking of $SL(16,C)$ to $%
SL(2,C)\times SU(8)$, can be written in the form

\begin{equation}
U=-M_{U}^{2}(\Phi ^{2}+\Phi ^{\prime 2})+h_{1}(\Phi ^{2}\Phi ^{2}+\Phi
^{\prime 2}\Phi ^{\prime 2})^{2}+h_{2}(\Phi ^{4}+\Phi ^{\prime
4})+h_{3}(\Phi ^{2}\Phi ^{\prime 2})  \label{u}
\end{equation}%
which is presumably induced by the multi-preon interactions at large
distances (here the notations $\Phi ^{2}=Tr(\Phi ^{+}\Phi )$, $\Phi ^{\prime
2}=Tr(\Phi ^{\prime +}\Phi ^{\prime })$, $\Phi ^{4}=Tr(\Phi ^{+}\Phi \Phi
^{+}\Phi )$ and $\Phi ^{\prime 4}=Tr(\Phi ^{\prime +}\Phi ^{\prime }\Phi
^{\prime +}\Phi ^{\prime })$ were used). Likewise, these interactions may
also produce, among others, the invariant Yukawa-type couplings for the
composite multiplets (\ref{216})
\begin{equation}
L_{Y}=\frac{1}{M_{Y}^{2}}\left( \overline{\Psi }_{Lr}^{[pq]}\widehat{D}\Psi
_{L[tu]}^{s}\Phi _{\lbrack pqs]}\Phi ^{\lbrack rtu]}+\overline{\Psi }%
_{Rr}^{[pq]}\widehat{D}\Psi _{R[tu]}^{s}\Phi _{\lbrack pqs]}^{\prime }\Phi
^{\prime \lbrack rtu]}\right)  \label{sup'}
\end{equation}%
\newline
($p,q,r,s,t,u=1,...,8$) being initially L-R symmetric as well. The mass
parameter $M_{Y}$ is some UV cutoff which in our case can be ultimately
related to the preon confinement energy scale $\Lambda _{MC}$.

As in the known left-right models \cite{moh}, in the potential (\ref{u}) for
a natural range of the parameters, particularly, for $h_{3}>2(h_{1}+h_{2})$,
and for properly chosen higher-dimensional coupling constants, the scalars $%
\Phi _{\lbrack ijk]}$ and $\Phi _{\lbrack ijk]}^{\prime }$ may develop the
totally antisymmetric VEV configuration
\begin{equation}
\left\langle \Phi ^{\lbrack pqr]}\right\rangle =0\text{ , \ }\left\langle
\Phi ^{\prime \lbrack pqr]}\right\rangle =\delta _{\overline{p}}^{p}\delta _{%
\overline{q}}^{q}\delta _{\overline{r}}^{r}\epsilon ^{\overline{p}\overline{q%
}\overline{r}}M_{LR}\text{ \ \ \ }(\overline{p},\overline{q},\overline{r}%
=1,2,3)  \label{ph}
\end{equation}%
which leads at the scale $M_{LR}$ to the spontaneous breakdown of the
starting $SU(8)$ down to the $SU(5)\times SU(3)$ symmetry group which we may
identify as the conventional $SU(5)$ grand unification times the $SU(3)_{F}$
family symmetry. Moreover, due to\ asymmetric VEV (\ref{ph}), the Yukawa
couplings (\ref{sup'}) certainly lead to the chiral symmetry breaking for
the right-handed composites, $SU(8)_{R}\rightarrow \lbrack SU(5)\times
SU(3)]_{R}$, leaving intact the $SU(8)_{L}$ symmetry for the left-handed
ones. In terms of the spectator gauge chiral symmetry, this means that all
non-diagonal gauge bosons related to the broken generators of the coset $%
SU(8)_{R}\ /$ $[SU(5)\times SU(3)]_{R}$ acquire at large distances the scale
$M_{LR}$ order masses. This means that, though the massless right-handed
preons still possess the $SU(8)_{R}$ symmetry, the masslessness of their
composites at large distances is now controlled solely by the remaining $%
[SU(5)\times SU(3)]_{R}$ part. Indeed, the right-handed composites no longer
fill the complete $216_{R}$ multiplet but instead satisfy a modified anomaly
matching condition based on $[SU(5)\times SU(3)]_{R}$. One finds that the
right-handed spectrum consistent with these anomalies excludes the multiplet
$(5+10,3)_{R}$, leaving only the submultiplet combination
\begin{equation}
\lbrack (45,1)+(5,8+1)+(5+\overline{5},3)+(1,\overline{6})]_{R}  \label{r}
\end{equation}%
which satisfies the $SU(5)^{3}$ and $SU(3)^{3}$ anomaly matching conditions
with the right-handed $Q$ preons. When the left-handed multiplet (\ref{216})
and the reduced right-handed multiplet (\ref{r}) are combined, all
components except $(\overline{5}+10,\overline{3})_{L}$ can pair up and
become heavy, leaving only
\begin{equation*}
(\overline{5}+10,\text{ }\overline{3})_{L}+(24+1,3)_{L}+(5+\overline{5}%
,3)_{R}
\end{equation*}%
as the anomaly-free light spectrum, with $(24+1,3)_{L}$ and $(5+\overline{5}%
,3)_{R}$ expected to be lifted to higher scales through the family-symmetry
interactions. Thus three chiral families of quarks and leptons naturally
remain light, transforming as $(\overline{5}+10,\overline{3})_{L}$.

Once L-R symmetry is broken in this way, the initially vectorlike metaflavor
symmetry $SU(8)\times SL(2,C)$ effectively reduces in the composite sector
to a chiral subgroup $[SU(5)\times SU(3)_{F}]\times SL(2,C)$, where $%
SU(3)_{F}$ is a family symmetry acting on the three quark-lepton generations
\cite{h}. Further symmetry breaking is then driven by scalar multiplets
(which may themselves be composite) that break $SU(5)$ and $SU(3)_{F}$ to
the Standard Model, generating the observed pattern of gauge bosons and
fermion masses and mixings.

\subsection{Down to the Standard Model}

A conventional scenario for further breaking of the $SU(5)\times SU(3)_{F}$
symmetry---identified here as the GUT and family symmetry---introduces a
suitable set of scalar fields that may reduce this theory down to the
Standard Model. For this, one takes the $SL(16,C)$ adjoint scalar multiplets
of the type
\begin{equation}
\Sigma =(\sigma ^{k}+i\sigma _{5}^{k}\gamma _{5})\lambda ^{k}+\sigma
_{ab}^{K}\gamma ^{ab}\lambda ^{K}/2  \label{ad}
\end{equation}%
($k=1,...,64;$ $K=0,k$) which transform as
\begin{equation*}
\Sigma \rightarrow \Omega \Sigma \Omega ^{-1}\text{\ }
\end{equation*}%
In general, $\Sigma $ contains not only scalar components but also
pseudoscalar and spinorial tensor components in (\ref{ad}), which can be
filtered out by the covariant tetrad-constraint mechanism. Imposing on $%
\Sigma $\ \ the constraint
\begin{equation*}
\Sigma =e_{\rho }\Sigma e^{\rho }/4
\end{equation*}%
one finds, after tetrad condensation, that $\Sigma $ is effectively
\textquotedblleft sandwiched\textquotedblright\ by the neutral tetrads\ (\ref%
{or6}). As a result, only the pure scalar components remain in the $SU(5)$
symmetry breaking multiplet, $\Sigma =\sigma ^{k}\lambda ^{k}$, thereby
enabling the breaking of the $SU(5)$ GUT down to the Standard Model, $%
SU(5)\rightarrow SU(3)_{c}\times SU(2)_{W}\times U(1)_{Y}$, caused by the
conventional potential \cite{moh}.

The final breaking of the SM and of the accompanying family symmetry $%
SU(3)_{F}$ to $SU(3)_{c}\times U(1)_{em}$ occurs through the extra
multiplets with assignments determined by representations chosen for quarks
and leptons. They are $H^{[pa,qb,rc,sd]}$, and $\xi _{\lbrack pa,qb]}$ and $%
\chi _{\{pa,qb\}}$ of $SL(16,C)$, which contain an even number of
antisymmetrized and symmetrized $SU(8)$ and $SL(2,C)$ indices ($%
p,q,r,s,t,u=1,...,8;$ $a,b,c,d,e,f=1,2$). These multiplets include, among
other components, genuine scalar components that develop the corresponding
VEVs giving masses to the weak bosons, as well as to the family bosons of
the $SU(3)_{F}$. They also generate masses for quarks and leptons located in
the left-handed fermion multiplet (\ref{216}) through the $SL(16,C)$
invariant Yukawa couplings. Particularly, for the submultiplet $(\overline{5}%
+10,\overline{3})_{L}$ one has, as in a conventional $SU(5)$ GUT, the two
independent couplings

\begin{eqnarray}
&&\frac{1}{\emph{M}}\left[ \Psi _{\lbrack qa,\text{ }rb]L}^{pc}C\Psi
_{\lbrack te,\text{ }uf]L}^{sd}\right] H^{\{[qa,rb],[te,uf]\}}\left(
a_{U}\xi _{\lbrack pc,sd]}+b_{U}\chi _{\{pc,sd\}}\right) ,  \notag \\
&&\frac{1}{\emph{M}}\left[ \Psi _{\lbrack qa,\text{ }rb]L}^{pc}C\Psi
_{\lbrack pc,\text{ }te]L}^{sd}\right] H^{\{[qa,rb],[te,uf]\}}(a_{D}\xi
_{\lbrack sd,uf]}+b_{D}\chi _{\{sd,uf\}})  \label{yc}
\end{eqnarray}
with distinct index contraction for the up quarks, and down quarks and
leptons, respectively. The mass $M$ represents an effective scale in the
theory that, in the composite model of quarks and leptons, can be linked to
their compositeness scale $\Lambda _{MC}$, while $a_{U,D}$ and $b_{U,D}$ are
dimensionless constants of the order of $1$.

Actually, these couplings contain two types of scalar-containing multiplets
with the following $SU(8)\times SL(2,C)$ components---the $H$ multiplet with
the scalar components
\begin{equation}
H^{[pqrs]\{[ab],[cd]\}}(70,1)  \label{70}
\end{equation}%
and the $\xi $ and $\chi $ multiplets, whose scalar components look as
\begin{equation}
\xi _{\lbrack pq\boldsymbol{]}[ab]}(28,1),\text{ \ }\chi _{\{pq\boldsymbol{\}%
}[ab]}(36,1)  \label{28}
\end{equation}%
Decomposing them into the components of the final $SU(5)\times SU(3)_{F}$
symmetry one finds the full set of scalars
\begin{eqnarray*}
70 &=&(5,1)+(\overline{5},1)+(10,\overline{3})+(\overline{10},3) \\
28 &=&(5,3)+(10,1)+(1,\overline{3}) \\
36 &=&(5,3)+(15,1)+(1,6)
\end{eqnarray*}%
containing the $SU(5)$ quintets $(5,1)$ and $(\overline{5},1)$ to break the
Standard Model at the electroweak scale $M_{SM}$, and the $SU(3)_{F}$
triplet and sextet, $(1,\overline{3})$ and $(1,6)$, to properly break the
family symmetry at some larger scale $M_{F}$. One may refer to the scalars ( %
\ref{70}) and (\ref{28}) as the corresponding "vertical" and "horizontal"
scalars, which provide the simplest form of the above Yukawa couplings.
Acting in pairs, they presumably determine the masses and mixings of all
quarks and leptons. Lastly, and importantly, in the model under
consideration, these scalars may themselves be composite states formed from
the same preons as quarks and leptons \cite{ch}.

After the $SU(8)$ symmetry breaking the Yukawa couplings (\ref{yc}) acquire
the transparent $SU(5)\times SU(3)_{F}$ invariant form (all metaflavor
indices are omitted)
\begin{eqnarray}
&&\left[ (10,\overline{3})_{L}C(10,\overline{3})_{L}\right]
(5,1)_{U}[a_{U}(1,\overline{3})+b_{U}(1,6)]/\mathcal{M}\text{ ,}  \notag \\
&&[(\overline{5},\overline{3})_{L}C(10,\overline{3})_{L}](\overline{5}%
,1)_{D}[a_{D}(1,\overline{3})+b_{D}(1,6)]/\mathcal{M}\text{ }  \label{ud'}
\end{eqnarray}%
where we only include those components of the vertical scalar $H$ and
horizontal scalars $\xi $ and $\chi $ which develop the VEVs. Just the
horizontal scalar VEVs determine through the Yukawa couplings (\ref{ud'})
the mass matrices for quarks and leptons
\begin{eqnarray*}
\widehat{m}_{\overline{p}\overline{q}}^{U} &=&\left\langle 5,1\right\rangle
_{U}[a_{U}\left\langle 1,\overline{3}\right\rangle _{[\overline{p}\overline{q%
}]}+b_{U}\left\langle 1,6\right\rangle _{\{\overline{p}\overline{q}\}}]/%
\mathcal{M}\text{ \ } \\
\widehat{m}_{\overline{p}\overline{q}}^{D} &=&\left\langle \overline{5}%
,1\right\rangle _{D}[a_{D}\left\langle 1,\overline{3}\right\rangle _{[%
\overline{p}\overline{q}]}+b_{D}\left\langle 1,6\right\rangle _{\{\overline{p%
}\overline{q}\}}]/\mathcal{M}\text{ \ }
\end{eqnarray*}%
where the angle brackets denote the corresponding VEVs, while $\overline{p},%
\overline{q}=1,2,3$ stand for family indices. The matrices $\widehat{m}_{%
\overline{p}\overline{q}}^{U}$ and $\widehat{m}_{\overline{p}\overline{q}%
}^{D}$ are defined at the grand unified scale and must then be extrapolated
down to the physical mass range of quarks and leptons. Depending on which
components of the symmetric and asymmetric VEVs develop, different
texture-zero patterns arise for the mass matrices. The strong hierarchies of
the quark-lepton masses and mixings may then be explained by softer
hierarchies among the breaking directions of the $SU(3)_{F}$ family
symmetry, whose scale $M_{F}$\ is assumed to lie close to the effective
scale $M$.

The composite nature of quarks and leptons introduces, in addition to the
Planck scale $M_{P}$ and the $SL(16,C)$ breaking scale $M$ (determined by
the tetrad condensate), a new fundamental scale: the preon confinement scale
$\Lambda _{MC}$. Thus the same $SL(16,C)$ metaflavor structure organizes
both the gauge-gravity sector and the preonic origin of the observed quarks
and leptons, while their composition is governed by the independent
metacolor dynamics at the scale $\Lambda _{MC}$. The phenomenology depends
on the relative hierarchy among these three scales. If $M$, $\Lambda _{MC}$
and $M_{P}$ are all near one another, the framework mainly provides a
structural reinterpretation of known physics. If $M$ and $\Lambda _{MC}$ are
considerably below $M_{P}$, a rich spectrum of preonic and hyperunified
states may become accessible, offering possible experimental windows into
the underlying $SL(16,C)$ hyperunification.


\section{Summary and outlook}

We have examined a class of hyperunified gauge theories based on the
non-compact group $SL(2N,C)$ that aim to unify gravity with internal
interactions. Gauging neutral spinors leads to $SL(2,C)$ as the natural
local symmetry associated with spacetime spin structure, while gauging
charged spinors extends the symmetry to $SL(2N,C)$, with an internal $SU(N)$
hyperflavor subgroup. Within this framework the accompanying tetrads play a central role: once treated as dynamical fields, the nonlinear length constraint fixes their soldering norm, while quantum effects select the hyperflavor-blind vacuum orientation; the resulting configuration filters out the non-compact directions in the gauge algebra and supports a radiative origin of Einstein--Cartan gravity.

The low-energy symmetry that remains after tetrad condensation through the
hyperflavor-blind vacuum is $SL(2,C)\times SU(N)$, realized on a spectrum in
which only the neutral tetrad, identified with the graviton, and the $SU(N)$
vector field multiplet are massless. Starting from a unified
quadratic-strength Lagrangian with a single gauge coupling, we argued that
the EC term can be induced by fermion loops involving the flavor-singlet
tensor connection and the tetrad, with either heavy vectorlike fermions near
the Planck scale or a universal UV cutoff providing the required scale. At
the same time, the Teleparallel-Gravity (TGR) and Ghost-Free-Combination
(GFC) structures, described in Sections 3 and 4, play the role of no-ghost
kinetic projectors: TGR selects the healthy massless spin-2 dynamics of the
neutral tetrad, while GFC ensures the admissible propagation of the
spin-connection sector.

A distinctive feature of the whole construction is that gravitational
unification does not appear as an ordinary exact high-symmetry phase
followed by conventional symmetry breaking, as in internal GUTs. The full $%
SL(2N,C)$ covariance is present in the premetric formulation, but the
physical meaning of gravity emerges only after tetrad soldering selects the
neutral spacetime branch. Thus the theory is not merely a
representation-level embedding of Lorentz and internal generators into a
larger algebra; it becomes a dynamical hyperunification only when the
gravitational scales are generated by the same gauge-coupled matter sector
rather than inserted as independent constants.

On the matter side, we saw that attempting to assign elementary quarks and
leptons to $SL(2N,C)$ multiplets usually leads to large representations with
unwanted spin content. This motivates a composite viewpoint: the truly
elementary spinor fields are not the observed quarks and leptons themselves,
but preons in fundamental $SL(2N,C)$ representations, while quarks and
leptons arise as their bound states. Applying the 't~Hooft anomaly matching
condition to the global chiral symmetry $SU(N)_{L}\times SU(N)_{R}$ of
preons, while assuming that only three-preon spin-$1/2$ composites remain
massless and fit into a single representation, singles out $SU(8)_{L}\times
SU(8)_{R}$ at large distances and thus $SL(16,C)$ as the hyperunified
metaflavor group. The residual symmetry $SL(2,C)\times SU(8)$ naturally
yields three families of composite quarks and leptons, organized into an $%
SU(5)\times SU(3)_{F}$ structure after L--R breaking in the composite sector.

Remarkably, compositeness appears here to be dictated by the structure of the theory itself, rather than introduced solely on phenomenological grounds such as the existence of identical quark-lepton families and their mass and mixing hierarchies, as is usually the case. In the present context, applying the fundamental $SL(2N,C)$ symmetry directly to the observed quarks and leptons would be somewhat analogous to imagining the Standard Model, or its conventional GUT extensions, being applied directly to hadrons in order to describe their classification and interactions. Such an identification would obscure the underlying constituent structure and could therefore be fundamentally misleading.

The proposed hyperunification mechanism is purely four-dimensional and
gauge-based; it does not rely on extra dimensions \cite{add,rs,kp} or string
excitations \cite{str}. The tetrad, which is required in any case to couple spinors to gravity, is used as the field that condenses; its dynamically selected hyperflavor-blind orientation fixes the vacuum and lifts non-compact directions with a clear understanding of which gauge components have been decoupled. In this way the gauge-gravity construction
and the preon construction meet at the level of the same $SL(16,C)$
metaflavor symmetry: the former uses it to organize the unified connection
and tetrad sector, while the latter uses it to classify the truly elementary
preons whose metacolor bound states become the observed quarks and leptons.

Further work should complete the preon model for $N=8$, including the full
set of composites. If the tetrad-condensation scale $M$ and the preon
confinement scale $\Lambda _{MC}$ lie well below $M_{P}$, remnants of the
filtered hyperunified structure could leave characteristic signatures in the
spectrum of heavy bosonic and fermionic composites. A related possibility is
that all $SL(2N,C)$ gauge fields themselves may be composite, built from
preon bilinears with only global $SL(2N,C)$ imposed at the fundamental
level. This approach, long recognized as a viable alternative to
conventional gauge theories and gravity \cite{bjorken, ph, eg, ter, suz,
jlc, jpl2}, has not yet been explored in the context of noncompact unified
symmetries. Extending the present work to such a fully composite setting
remains an interesting direction for future study.

\begin{acknowledgments}
I am grateful to Colin Froggatt, Oleg Kancheli, Holger Nielsen and Grigory Volovik for illuminating discussions, and Lin Adams for reading the manuscript and useful remarks.
\end{acknowledgments}


\begin{thebibliography}{99}
\bibitem{Utiyama} R. Utiyama, Phys. Rev. \textbf{101}, 1597 (1956).

\bibitem{Kibble} T. W. Kibble, J. Math. Phys. \textbf{2}, 212 (1960).

\bibitem{YM} C. N. Yang and R. L. Mills, Phys. Rev. \textbf{96}, 191 (1954).

\bibitem{ish} C. J. Isham, A. Salam, and J. Strathdee, Lett. Nuovo Cim. \textbf{5}, 969 (1972).

\bibitem{ish1} C. J. Isham, A. Salam, and J. Strathdee, Phys. Rev. D \textbf{8}, 2600 (1973); Phys. Rev. D \textbf{9}, 1702 (1974).

\bibitem{cho} Y. M. Cho and P. G. O. Freund, Phys. Rev. D \textbf{12}, 1711 (1975); Y. M. Cho, Phys. Rev. D \textbf{14}, 3335 (1976).

\bibitem{hu} J. C. Huang and P. W. Dennis, Phys. Rev. D \textbf{15}, 983 (1977); Phys. Rev. D \textbf{24}, 3125 (1981).

\bibitem{per} R. Percacci, Phys. Lett. B \textbf{144}, 37 (1984); F. Nesti and R. Percacci, J. Phys. A \textbf{41}, 075405 (2008); F. Nesti and R. Percacci, Phys. Rev. D \textbf{81}, 025010 (2010).

\bibitem{cham} A. H. Chamseddine, Phys. Rev. D \textbf{70}, 084006 (2004).

\bibitem{jpl} J. L. Chkareuli, Phys. Lett. B \textbf{834}, 137417 (2022); Eur. Phys. J. C \textbf{84}, 1212 (2024); Nucl. Phys. B \textbf{1020}, 117150 (2025).

\bibitem{18} S. Coleman and J. Mandula, Phys. Rev. \textbf{159}, 1251 (1967).

\bibitem{tg} K. Hayashi and T. Shirafuji, Phys. Rev. D \textbf{19}, 3524 (1979); R. Aldrovandi and J. G. Pereira, \textit{Teleparallel Gravity: An Introduction, Fundamental Theories of Physics} (Springer, 2013).

\bibitem{W} Steven Weinberg, \textit{The Quantum Theory of Fields, Volume II: Modern Applications} (Cambridge University Press, 1996).
\bibitem{nev} D. E. Neville, Phys. Rev. D \textbf{18}, 3535 (1978).

\bibitem{nev1} E. Sezgin and P. van Nieuwenhuizen, Phys. Rev. D \textbf{21}, 3269 (1980).

\bibitem{heh} F. W. Hehl, P. von der Heyde, G. D. Kerlick, and J. M. Nester, Rev. Mod. Phys. \textbf{48}, 393 (1976).

\bibitem{cw} S. Coleman and E. Weinberg, Phys. Rev. D \textbf{7}, 1888 (1973).

\bibitem{pdg} S. Navas \textit{et al.} (Particle Data Group), Phys. Rev. D \textbf{110}, 030001 (2024).

\bibitem{Sakharov:1967} A. D. Sakharov, Sov. Phys. Dokl. \textbf{12}, 1040 (1968).

\bibitem{Adler:1982ri} S. L. Adler, Rev. Mod. Phys. \textbf{54}, 729 (1982).

\bibitem{Zee:1981} A. Zee, Phys. Rev. D \textbf{23}, 858 (1981).

\bibitem{Don} J. F. Donoghue, Phys. Rev. D \textbf{50}, 3874 (1994).

\bibitem{gg} H. Georgi and S. L. Glashow, Phys. Rev. Lett. \textbf{32}, 438 (1974).

\bibitem{su} S. Dimopoulos and H. Georgi, Nucl. Phys. B \textbf{193}, 150 (1981).

\bibitem{98} J. Hisano, H. Murayama, and T. Yanagida, Nucl. Phys. B \textbf{402}, 46 (1993).

\bibitem{99} J. L. Chkareuli and I. G. Gogoladze, Phys. Rev. D \textbf{58}, 055011 (1998).

\bibitem{ge} H. Georgi, Nucl. Phys. B \textbf{156}, 126 (1979).

\bibitem{ch} J. L. Chkareuli, Nucl. Phys. B \textbf{941}, 425 (2019); Nucl. Phys. B \textbf{1020}, 117150 (2025).

\bibitem{t} G. 't Hooft, in \textit{Recent Developments in Gauge Theories}, edited by G. 't Hooft \textit{et al.} (Plenum, New York, 1980).

\bibitem{h} J. L. Chkareuli, JETP Lett. \textbf{32}, 671 (1980); Z. G. Berezhiani and J. L. Chkareuli, Sov. J. Nucl. Phys. \textbf{37}, 618 (1983); J. L. Chkareuli, C. D. Froggatt, and H. B. Nielsen, Nucl. Phys. B \textbf{626}, 307 (2002).

\bibitem{moh} R. N. Mohapatra, \textit{Unification and Supersymmetry} (Springer-Verlag, New York, 2003).

\bibitem{add} I. Antoniadis, N. Arkani-Hamed, S. Dimopoulos, and G. Dvali, Phys. Lett. B \textbf{436}, 257 (1998).

\bibitem{rs} L. Randall and R. Sundrum, Phys. Rev. Lett. \textbf{83}, 3370 (1999).

\bibitem{kp} K. Krasnov and R. Percacci, Class. Quantum Grav. \textbf{35}, 143001 (2018).

\bibitem{str} K. Becker, M. Becker, and J. H. Schwarz, \textit{String Theory and M-Theory: A Modern Introduction} (Cambridge University Press, 2007).

\bibitem{bjorken} J. D. Bjorken, Ann. Phys. (N.Y.) \textbf{24}, 174 (1963).

\bibitem{ph} P. R. Phillips, Phys. Rev. \textbf{146}, 966 (1966).

\bibitem{eg} T. Eguchi, Phys. Rev. D \textbf{14}, 2755 (1976).

\bibitem{ter} H. Terazawa, Y. Chikashige, and K. Akama, Phys. Rev. D \textbf{15}, 480 (1977).

\bibitem{suz} M. Suzuki, Phys. Rev. D \textbf{37}, 210 (1988); Phys. Rev. D \textbf{82}, 045026 (2010).

\bibitem{jlc} J. L. Chkareuli, C. D. Froggatt, and H. B. Nielsen, Phys. Rev. Lett. \textbf{87}, 091601 (2001).

\bibitem{jpl2} J. L. Chkareuli, Phys. Lett. B \textbf{817}, 136281 (2021).
\end{thebibliography}
\end{document}